\documentclass[11pt,reqno]{article}
\usepackage{amsmath}
\usepackage{amsfonts}
\usepackage{amssymb}
\usepackage{latexsym}
\usepackage[dvips]{graphicx}
\usepackage{epsf}

\allowdisplaybreaks \numberwithin{equation}{section}

\def \ee{\end{equation}}
\def \be{\begin{equation}}
\def \bea{\begin{eqnarray}}
\def \eea{\end{eqnarray}}

\newcommand{\cC}{\mathcal{C}}
\newcommand{\cD}{\mathcal{D}}

\newcommand{\cM}{\mathcal{M}}

\newcommand{\cR}{\mathcal{R}}

\newcommand{\m}{\mu}
\newcommand{\n}{\nu}
\newcommand{\p}{\partial}
\newcommand{\half}{\tfrac{1}{2}}
\newcommand{\gb}{\bar{g}}

\begin{document}

%
%

\bigskip
\bigskip
\bigskip
\thispagestyle{empty}

\renewcommand*{\thefootnote}{\fnsymbol{footnote}}
\begin{center}
 {\LARGE\bfseries  Black holes within Asymptotic Safety\footnote{Invited review for International Journal of Modern Physics A.}}
\\[10mm]
Benjamin Koch$^1$ and Frank Saueressig$^2$ \\[3mm]
{\small\slshape
$^1$ Instituto de F\'isica, Pontificia Universidad Cat\'olica de Chile \\
Av. Vicu\~na Mackenna 4860\\
Santiago, Chile \\
{\upshape\ttfamily bkoch@fis.puc.cl}} \\[1.1ex]
{\small\slshape
$^2$ Institute for Mathematics, Astrophysics and Particle Physics 
(IMAPP) \\
Radboud University Nijmegen, Heyendaalseweg 135\\
 6525 AJ Nijmegen, The Netherlands\\ 
{\upshape\ttfamily f.saueressig@science.ru.nl}} \\[1.1ex]
\end{center}
\vspace{8mm}


\hrule\bigskip

\centerline{\bfseries Abstract} \medskip
\noindent
Black holes are probably among the most fascinating objects populating our universe. Their characteristic features found within general relativity, encompassing spacetime singularities, event horizons, 
and black hole thermodynamics, provide a rich testing ground for quantum gravity ideas.  We review the status of black holes within a particular proposal for quantum gravity, Weinberg's asymptotic safety program. Starting from a brief survey of the effective average action and scale setting procedures, 
an improved quantum picture of the black hole is developed. The Schwarzschild black hole and its generalizations including angular momenta, higher-derivative corrections and the implications of extra dimensions are discussed in detail. In addition, the quantum singularity emerging for the inclusion of a cosmological constant is elucidated and linked to the phenomenon of a dynamical dimensional reduction of spacetime.
\bigskip
\hrule\bigskip
\renewcommand*{\thefootnote}{\arabic{footnote}}
\newpage


\tableofcontents

\section{Introduction}
Black holes \cite{Carroll:2004st,Wald:1995yp,Taylor,Poisson} are fascinating objects, which 
are in the focus of many scientific communities  
including observational astrophysicists, general relativists, and theoreticians working on quantum 
gravity. By definition, a black hole is a region of spacetime from which gravity prevents anything, including light, from escaping. From
the observational side there is hardly any doubt that a supermassive black hole with a mass of order $2 \times 10^6$ solar masses sits in the center
of our galaxy \cite{schroeder:nature}. Moreover, there is an active search for primordial black holes which could have formed in the very early universe
and enter into the final stage of their lifetime today \cite{Carr:2009jm}. Microscopic black holes have been advocated as candidates for massive astrophysical compact halo objects (MACHOs) that could 
account for the apparent presence of dark matter. 
These experimental searches go hand in hand with the effort of the theoretical physics community to  
develop the theoretical understanding of these objects, since 
they provide an important laboratory for testing ideas related to quantum
gravity. The goal of the present article
is to provide a status report summarizing the understanding of black hole physics within the gravitational asymptotic safety program
also called Quantum Einstein Gravity (QEG) \cite{Niedermaier:2006wt,Reuter:2007rv,Percacci:2007sz,Litim:2008tt,Reuter:2012id}.
For similar accounts in string theory and loop quantum gravity we refer to Ref.\ \cite{Bena:2013dka} and Ref.\ \cite{FrancescaBook}, respectively.

In classical general relativity black holes appear as vacuum solutions of Einstein's equations. The simplest
black hole solution is described by the Schwarzschild metric
\be\label{ssmetric}
ds^2 = -f(r) \, dt^2+ f(r)^{-1} \, dr^2 + r^2 d\Omega_2^2
\ee
with $d\Omega_2^2$ denoting the line-element of the two-sphere and the radial function
\be\label{frfct}
f(r) = 1 - \frac{2 G M}{r} \, . 
\ee
By Birkhoff's theorem, the Schwarzschild metric constitutes the 
unique spherically symmetric vacuum solution. Besides depending on Newton's constant $G$ the solution is completely
characterized by the mass $M$ of the black hole. Later on this solution
was systematically generalized to include angular momentum $J$ (Kerr metric), electric charges $Q$ (Reissner-Nordstr\"om metric),
and a combination of the two (Kerr-Newman metric). Generalizations to spacetimes with more and less than four dimensions
have led to the discovery of the BTZ black hole, Ref.\ \cite{Banados:1992wn}, and revealed the breakdown of the 
 uniqueness theorems once the spacetime includes extra dimensions \cite{Emparan:2001wn}. 
 
As a characteristic feature, classical black holes possess a gravitational singularity where curvature invariants like the square
of the Riemann tensor diverge. Typically, these singularities are shielded by an event horizon.\footnote{A solution of Einstein's equations
giving rise to a naked singularity (i.e., a singularity not shielded by an event horizon) is the Kerr black hole with $J > G M^2$.} This feature can be formalized in terms of the cosmic censorship hypothesis stating that the gravitational singularity is hidden from an observer at infinity. The occurrence of singularities in general relativity is often interpreted as a signal that the theory is incomplete and a consistent description requires a treatment within a (hitherto elusive) quantum theory of gravity. Vice versa, the capability of resolving the classical black hole singularity is often advocated as a benchmark test for the viability of the quantum theory. 

The first milestone towards a quantum treatment of black holes was the discovery that black holes emit Hawking radiation and are thus not entirely black \cite{Hawking:1974sw}. 
One way to obtain this result is by quantizing a scalar field in
a black hole background, leading to the prediction that the black hole emits black body radiation with temperature inversely proportional to its mass. More generally,
given a black hole or cosmological horizon situated at $r_{\rm H}$ the temperature $T_{\rm H}$ of the black body radiation emitted by the horizon is
\be\label{HT}
T_{\rm H} = \frac{1}{4\pi} \, \left. \p_r f(r) \right|_{r = r_{\rm H}} \, . 
\ee
 Heuristically,
the origin of this thermal radiation can be understood in terms of pair production of particles close to the event horizon. One partner crosses the horizon and falls into the singularity, while the other becomes a real particle escaping to infinity where it is detected as black hole radiation. The particle falling towards the singularity thereby generates a negative energy flow into the black hole so that its mass decreases. This constituted an important step towards developing the thermodynamics of black holes, which associates an entropy $S_{\rm BH}$ to the event horizon. Semi-classically, $S_{\rm BH}$ is given by the Bekenstein-Hawking area law, 
\be\label{BHent}
S_{\rm BH} = \frac{A}{4 \, G} \, , 
\ee
where $A$ is the area of the event horizon. Subleading corrections to this formula have been computed in
many quantum gravity proposals \cite{Kiefer:2012boa}.

In this review, we will focus on the role of black holes within a specific 
proposal for a quantum theory of gravity, the gravitational asymptotic safety program \cite{Niedermaier:2006wt,Reuter:2007rv,Percacci:2007sz,Litim:2008tt,Reuter:2012id}.
This scenario, first advocated by Weinberg\cite{Weinberg:1979}, and inaugurated by the introduction of functional renormalization group (RG) methods by Reuter
in Ref.\ \cite{mr}, proposes that gravity is a nonperturbatively renormalizable quantum field theory. The key ingredient of the scenario is a non-Gaussian fixed point (NGFP) of the theory's RG flow which controls the scale dependence of the coupling constants in the ultraviolet (UV).
The running of the coupling constants is thus special in its UV limit in the sense that all dimensionless combinations remain finite. This suffices to render physical quantities safe from unphysical divergences. Provided that the set of RG trajectories approaching the fixed point in the UV is finite-dimensional, i.e., 
parameterized by a finite number of ``relevant'' couplings, asymptotic safety is as predictive as a standard perturbatively renormalizable quantum field theory. 
While a rigorous existence proof for the NGFP is still lacking, there is by now substantial evidence supporting the viability of the asymptotic safety scenario \cite{Reuter:2012id}.

The key ingredient in developing the asymptotic safety program is the gravitational effective average action $\Gamma_k$ \cite{mr}.
By construction, $\Gamma_k$ contains effective vertices which include the effect of the quantum fluctuations with momenta $p^2 > k^2$.  
In this sense $\Gamma_k$ provides \emph{an effective description of physics} at the momentum scale $k^2 \approx p^2$.
This feature has been essential for assessing the phenomenological consequences of asymptotic safety.
Starting from the pioneering papers by Bonanno and Reuter, Refs.\ \cite{Bonanno:1998ye,Bonanno:2000ep},
the quantum properties of black holes arising from asymptotic safety have been explored systematically
by constructing RG improved solutions. Following up on these initial works, the dynamics of black hole evaporation based on the Vaidya-metric has been studied in Ref.\ \cite{Bonanno:2006eu},
while the RG improved Kerr metric has been constructed in \cite{Reuter:2006rg,Reuter:2010xb}. 
In \cite{Falls:2012nd} the results of \cite{Bonanno:2000ep} have been reanalyzed from a thermodynamics perspective
and the role of higher derivative terms in the effective average action has been explored in \cite{Cai:2010zh}. 
 Most recently RG improvements including scale-dependent surface terms have been carried out  by Becker and Reuter, Refs.\ \cite{Becker:2012js,Becker:2012jx},
and an equation highlighting the state-counting properties of $\Gamma_k$ has been proposed in Ref.\ \cite{Becker:2012js}.
Moreover, Refs.\ \cite{Koch:2013owa,Koch:2013rwa} established that, rather counter-intuitively, the cosmological constant plays a crucial role
in determining the short distance physics of the quantum black holes.
Further aspects of RG improved black hole geometries have been explored in Refs.\ \cite{Emoto:2005te,Ward:2006vw,Falls:2010he,Basu:2010nf,Casadio:2010fw}.
\footnote{Along similar lines the study of RG-improved cosmologies has been initiated in Refs.\ \cite{Bonanno:2001xi,Bonanno:2001hi}, recently reviewed in \cite{Reuter:2012xf}, while
RG-improved diffusion processes characterizing the microscopic structure of the asymptotically safe quantum spacetime
have been investigated in \cite{Lauscher:2005qz,Reuter:2011ah,Rechenberger:2012pm,Calcagni:2013vsa}.}
In general, these works established that quantum corrections in the vicinity of the nontrivial UV fixed points lead to drastic modifications of the classical picture at
short distances or high momenta while the long-distance properties of the improved black hole spacetime essentially agree with classical general relativity.

The rest of this review is organized as follows. In section \ref{sect.2} we introduce the properties of the effective average action
together with the most common scale setting procedures. This background provides the starting point for exploring physics applications of
asymptotic safety. Sections \ref{sect.3} and \ref{sect.4} review the features of the 
renormalization group improved black hole solutions in asymptotically flat space and
in the presence of a cosmological constant, respectively. The new material contained
in section \ref{sect.5} links the latter to the dynamical dimensional reduction
of the asymptotically safe quantum spacetime \cite{Lauscher:2001ya,Lauscher:2002sq,Lauscher:2005qz,Reuter:2011ah}.
A brief summary of further developments is given in section \ref{sect.7} and we conclude with an outlook in section \ref{sect.6}.

\section{The effective average action and scale setting procedures}
\label{sect.2}
In this section, we introduce the basic ingredients underlying
the quantum description of black holes within asymptotic safety.
\subsection{A primer to asymptotic safety}
\label{secPrimer}
Before being able to discuss the quantum nature of a black hole,
one needs to specify the input from the ``fundamental theory'' in which
the black hole will be studied. In the  asymptotic safety program
this input is provided by the gravitational effective average action $\Gamma_k$, Ref.\ \cite{mr}, 
a Wilson type effective action which provides an effective description of physics at the momentum scale $k$.
 More formally, the effective vertices obtained from $\Gamma_k$ already contain information
about loop corrections where quantum fluctuations with momenta $p^2 \ge k^2$ 
have been integrated out. In the spirit of the effective action, the quantum properties
of the system can then be studied within a tree-level analysis based on $\Gamma_k$.

The key property of $\Gamma_k$ is that its $k$-dependence is governed by the 
formally exact functional renormalization group equation (FRGE)
 \be\label{FRGE}
\p_k \Gamma_k[\Phi, \bar{\Phi}] = \frac{1}{2}  {\rm Tr}  \left[ \left( \Gamma_k^{(2)} + \cR_k \right)^{-1} \, \p_k \cR_k \right] \, .
\ee
 Here $\Gamma_k^{(2)}$ is the second functional derivative of $\Gamma_k$ with respect to the quantum fields $\Phi$ at fixed background fields $\bar{\Phi}$. The mode suppression operator $\mathcal{R}_k[\bar{\Phi}]$ provides a $k$-dependent mass-term for fluctuations with covariant momenta $p^2 \ll k^2$ and vanishes for $p^2 \gg k^2$. Its appearance in the numerator and denominator renders the trace (\mbox{Tr}) both infrared and UV finite with the main contribution coming from quantum fluctuations with momentum $p^2 \simeq k^2$. The FRGE is an exact equation without any perturbative approximations. Given an initial condition it determines $\Gamma_k$ for all scales uniquely. Its solutions interpolate between the bare (microscopic) action at $k \rightarrow \infty$ and the effective action $\Gamma[\Phi] = \Gamma_{k=0}\big[\Phi, \bar{\Phi}=\Phi\big]$ at $k \rightarrow 0$, provided that these limits exist (also see \cite{Manrique:2008zw} for a more detailed discussion). 

While obtaining exact solutions of the FRGE is notoriously hard, the flow equation
permits several approximation schemes. One is, of course, perturbation
theory, i.e., the expansion of the FRGE in a small coupling constant or in $\hbar$.
The detailed connection between RG flows obtained from the FRGE and the MSbar scheme has recently been
worked out \cite{Codello:2013bra}. The main strength of the FRGE is, however, that it also allows to extract
non-perturbative information in a rather systematic way.\footnote{Here ``non-perturbative'' should be understood as an approximation
that does not resort to an expansion in a small parameter.} Typical strategies
involve a derivative expansion, truncating $\Gamma_k$ to the interaction operators with the lowest mass dimension,
or a vertex expansion. Over the years, these techniques have provided substantial insights on the structure of the gravitational RG flow,
in particular supporting the existence of a non-trivial RG fixed point suitable for
Weinberg's Asymptotic Safety scenario (see Refs.\ \cite{Dou:1997fg,Souma:1999at,Lauscher:2001rz,Reuter:2001ag,Codello:2007bd,Machado:2007ea,Benedetti:2009rx,Niedermaier:2009zz,Eichhorn:2009ah,Nagy:2012rn,Codello:2013wxa,Benedetti:2012dx,Demmel:2012ub,Dietz:2012ic} for a selective list of original works). Moreover, it has been demonstrated that the RG flow emanating from this NGFP can continuously be connected to a classical regime where  general relativity provides a good approximation 
 for $\Gamma_k$ over a wide range of momentum scales\cite{Reuter:2001ag,Reuter:2004nx,Rechenberger:2012pm,Christiansen:2012rx}.

\subsubsection*{The Einstein-Hilbert truncation}
The simplest non-perturbative computation, called the single-metric Einstein-Hilbert truncation, approximates the
gravitational part of $\Gamma_k$ by the Euclidean Einstein-Hilbert
action
\be\label{action}
 \Gamma_k^{\rm grav}[g] = \frac{1}{16 \pi G_k} \int d^dx \sqrt{g} \left[-R + 2 \Lambda_k \right] \, ,  
 \ee
supplemented by standard gauge-fixing and ghost terms. Substituting this ansatz
into the FRGE \eqref{FRGE} and projecting the flow onto the volume and curvature terms contained in the ansatz
allows to read off the beta functions governing the scale dependence of Newton's constant $G_k$ and the cosmological constant $\Lambda_k$. 
The result is conveniently expressed in terms of the dimensionless coupling constants
\be\label{Gvong}
 g_k= G_k \, k^{d-2}\quad, \quad
\lambda_k= \Lambda_k \, k^{-2} \quad \, , 
\ee 
and has first been derived in Ref.\ \cite{mr}
\be\label{betaeq}
k \p_k g_k = \beta_g(g_k, \lambda_k) \, , \qquad k \p_k \lambda_k = \beta_\lambda(g_k, \lambda_k) \, ,
\ee
where
\be\label{betafcts}
\begin{split}
 \beta_\lambda(g, \lambda) = & \, ( \eta_N - 2) \lambda
+ \half \left(4\pi\right)^{1-d/2} g  \\
& \times \left[ 2d(d+1) \Phi^1_{d/2}(- 2 \lambda) - 8d \Phi^1_{d/2}(0) - d (d+1) \eta_N \tilde{\Phi}^1_{d/2}(-2 \lambda) \right] \, , \\
\beta_g(g, \lambda) = & \, (d-2+\eta_N) g \, . 
\end{split}
\ee
Here the anomalous dimension of Newton's constant $\eta_N$ is given by
\be\label{G14}
\eta_N(g, \lambda) = \frac{g B_1(\lambda)}{1 - g B_2(\lambda)}
\ee
with the following functions of the dimensionless cosmological constant:
\be\label{Bns}
\begin{split}
B_1(\lambda) \equiv & \,  \tfrac{1}{3} \left( 4 \pi \right)^{1-d/2}  \Big[
d(d+1)\Phi^1_{d/2-1}(-2\lambda) - 6 d (d-1) \Phi^2_{d/2}(-2\lambda)    \\
& \qquad \qquad \qquad - 4 d \Phi^1_{d/2-1}(0) - 24 \Phi^2_{d/2}(0) \Big] \, ,\\
B_2(\lambda) \equiv & \, - \tfrac{1}{6} (4 \pi)^{1-d/2} \left[d(d+1) \tilde{\Phi}^1_{d/2-1}(-2\lambda) - 6 d (d-1) \tilde{\Phi}^2_{d/2}(-2 \lambda) \right] \, . 
\end{split}
\ee
The threshold functions $\Phi^p_n(w)$ and $\tilde{\Phi}^p_n(w)$ encode 
the dependence of the beta functions on the coarse-graining 
operator $\cR_k$. 
For practical computation, it is necessary to specify $\cR_k$. A convenient choice
is the optimized cutoff, Ref.\ \cite{Litim:2001up}, for which the integrals appearing in the threshold functions
can be carried out analytically 
\be\label{phiopt}
\Phi^{{\rm opt;p}}_n(w) = \frac{1}{\Gamma(n+1)} \frac{1}{(1+w)^p} \, , \qquad 
\widetilde{\Phi}^{{\rm opt;p}}_n(w) = \frac{1}{\Gamma(n+2)} \frac{1}{(1+w)^p} \, . 
\ee
The system (\ref{betaeq}) constitutes a two-dimensional projection of the RG flow.
Inspecting \eqref{G14}, the anomalous dimension receives contributions from arbitrary powers of Newton's constant. In this sense the result goes beyond a perturbative loop-expansion in $G$.

\begin{figure}[t]
\centering
\includegraphics[width=10cm]{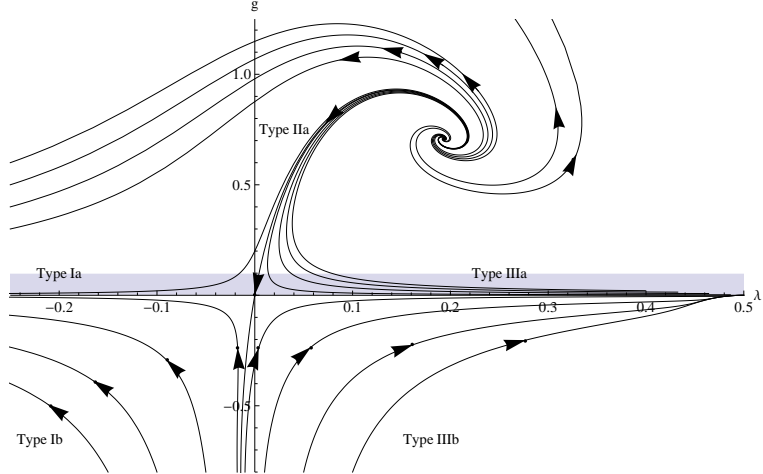}
\caption{\label{EHflow} Phase diagram obtained from integrating the beta functions of the Einstein-Hilbert truncation \eqref{betaeq} evaluated with the optimized regulator \eqref{phiopt}. The arrows point in the direction of increasing 
coarse-graining, i.e.\ of decreasing $k$. In the shaded region the dimensionful couplings $G_k$ and $\Lambda_k$ become scale-independent, so that classical general relativity emerges dynamically in the IR\cite{Reuter:2001ag}.}
\end{figure}
The scale dependence of $g_k, \lambda_k$ can be obtained by integrating 
the system \eqref{betaeq}. For $d=4$ and the optimized cutoff, the resulting phase portrait is shown in Fig.\ \ref{EHflow}. For positive Newton's constant the flow
it is dominated by the interplay of the Gaussian Fixed Point (GFP) situated in the origin $(g_* = 0, \lambda_* = 0)$ and the NGFP at
\be\label{NGFP}
\lambda_*=0.193 \, , \qquad g_*=0.707 \, , \qquad g_* \lambda_* = 0.137 \, .
\ee
Linearizing the flow around the NGFP, the eigenvalues
of the stability matrix $B_{ij} \equiv \p_{g_j} \beta_{g_i}|_{g = g_*}$
reveals that the NGFP is UV attractive for both $g_k, \lambda_k$.
Its stability coefficients, defined as minus the eigenvalues of $B_{ij}$
are
\be
\theta_{1,2} = 1.475 \pm 3.043 i \, . 
\ee
Notably, the existence and the stability properties of this NGFP are robust
with respect to changing the gauge-fixing conditions or the shape of $\cR_k$,
supporting the validity of the asymptotic safety scenario.

The NGFP acts as an UV attractor for all RG trajectories shown in the upper half-plane. Thus it governs the behavior of gravity at high energies and provides its UV-completion. 
Since the dimensionless couplings approach constant values, the NGFP also fixes the scaling behavior of the dimensionful Newton's constant and cosmological constant
at high energies
%
\be\label{UVG}
\lim_{k \rightarrow \infty} G_k = g_* \, k^{-2} \, , \qquad 
\lim_{k \rightarrow \infty} \Lambda_k = \lambda_* \, k^2\quad.
\ee
Following the RG flow towards the IR there is a crossover. Depending on whether the flow passes to the left, right, or ends at the GFP, the low-energy limit is given by classical general relativity with a negative (Type Ia), positive (Type IIIa) or zero (Type IIa, separatrix) value of the cosmological constant. The shaded region thereby illustrates the part of the phase diagram where the \emph{dimensionful} couplings are scale-independent, so that classical general relativity provides a good approximation of the physics in this region.

A consequence of the flow pattern shown in figure \ref{EHflow} and implied by the fixed point scaling \eqref{UVG} is the decrease of the \emph{dimensionful} Newton's constant for increasing $k$. This lends itself to the interpretation that gravity is anti-screening. The energy and momentum of the virtual particles surrounding every massive body experiences a gravitational pull towards this body, adding positively to its bare mass. Thus the cloud of virtual particles leads to an effective mass, that increases with increasing distance \cite{mr}.

\subsubsection*{Extension of the Einstein-Hilbert truncation by surface terms}
In view of black hole physics, an important generalization of the Einstein-Hilbert truncation is its extension
to spacetimes $\cM$ with boundary $\p \cM \not = 0$. This extension has been carried out in Ref.\ \cite{Becker:2012js}
and supplements the ``bulk'' Einstein-Hilbert action \eqref{action} with 
the Gibbons-Hawking boundary term
\be\label{GHterm}
\Gamma_k^\p = - \frac{1}{16 \pi G_k^\p} \int_{\p \cM} \, d^{d-1}x \, \sqrt{H} \left(2 K - 2 \Lambda_k^\p \right) \, . 
\ee
Here $H_{\m\n} = g_{\m\n} - n_\mu n_{\nu}$ is the boundary metric induced by $g_{\m\n}$, $n^\mu$ is the outward unit normal to the
surface $\p \cM$, and $K \equiv g^{\m\n} D_\m n_\n$ denotes the trace of the intrinsic curvature tensor.
 Besides the bulk Newton's constant $G_k$ and cosmological constant $\Lambda_k$, this ansatz contains
 two additional running coupling constants, the boundary Newton's constant $G_k^\p$ and a boundary cosmological constant 
 $\Lambda_k^\p$. For $G_k = G_k^\p$ the Einstein-Hilbert and Gibbons-Hawking term have the correct normalization
 for a well-posed variational problem, while the cosmological constants bear no special relationship.
A straightforward dimensional analysis shows that $G_k$ and $G_k^\p$ come with
the same canonical mass dimension while the dimensions of the cosmological constants
differ by one, $\left[ \Lambda^\p_k \right] = + 1$. Following \eqref{Gvong}, we introduce the dimensionless counterparts of the boundary coupling constants
\be\label{dimbound}
g_k^\p \equiv k^{d-2} \, G_k^\p \, , \qquad \lambda_k^\p \equiv k^{-1} \, \Lambda_k^\p \, . 
\ee
The couplings $\{g, g^\p, \lambda, \lambda^\p \}$ provide coordinates for the four-dimensional
theory space spanned by this truncation.

The beta functions arising from this ansatz have a block-diagonal form.
The flow of the bulk constants is independent of $g_k^\p, \lambda_k^\p$ and again
given by \eqref{betafcts}. This system is supplemented by the beta functions
for the boundary couplings
\be\label{betasurface}
k \p_k g_k^\p = \beta_{g^\p}(g_k, g_k^\p, \lambda_k) \, , \qquad k \p_k \lambda_k^\p = \beta_{\lambda^\p}(g_k, g_k^\p, \lambda_k, \lambda_k^\p) \, ,
\ee
where
\be
\begin{split}
\beta_{g^\p} = & \, \left( d-2 + \eta_N^\p \right) \, g^\p \, , \\
\beta_{\lambda^\p} = & \, \left( \eta_N^\p - 1 \right) \lambda^\p
- \frac{g^\p}{8 (4 \pi)^{(d-3)/2}} \left(2d(d+1) \Phi^1_{(d-1)/2}(-2\lambda) - 8d \Phi^1_{(d-1)/2}(0) \right. \\
& \qquad \qquad \qquad \qquad \qquad \qquad \qquad \left. - d(d+1) \eta_N \tilde{\Phi}^1_{(d-1)/2}(-2\lambda) \right) \, . 
\end{split}
\ee
Here 
\be
\begin{split}
\eta_N^\p(g, g^\p, \lambda) = \frac{g^\p}{3 (4 \pi)^{d/2-1}} & \,  \left(d(d+1) \Phi^1_{d/2-1}(-2\lambda) - 4d \Phi^1_{d/2-1}(0) \right. \\
& \; \; \left. - \tfrac{1}{2} d(d+1) \eta_N \tilde{\Phi}^1_{d/2-1}(-2\lambda)
\right) \, , 
\end{split}
\ee
is the anomalous dimension of the  boundary Newton's constant
and the anomalous dimension of the bulk Newton's constant $\eta_N(g, \lambda)$ is given by \eqref{G14}.

Owed to the block-diagonal structure, the phase diagram and in particular the fixed point structure of the bulk couplings
remains unaltered. Extending the analysis to the full system, one finds that the boundary sector also possesses a Gaussian (G) and non-Gaussian (NG) fixed point.
The fixed point structure is then given by the tensor product of these fixed points with their bulk counterparts so that one obtains the four fixed points
displayed in table \ref{b.fix}.
\begin{table}[t]
\begin{center}
\begin{tabular}{|c||c|c|c|c|c|}
 & $g_*$ & $\lambda_*$ & $g_*^\p$ & $\lambda_*^\p$ \\ \hline
 G-G-FP & $0$ & $0$ & $0$ & $0$ \\
 G-NG-FP & $0$ & $0$ & $-6\pi$ & $4/3$ \\
 NG-G-FP & $g_*$ & $\lambda_*$ & $0$ & $0$ \\
\; NG-NG-FP \; & \; $g_*$ \; & \; $\lambda_*$\; &\; $-12 \pi \frac{1-2\lambda_*}{7+16 \lambda_*}$\; &\; $\frac{4}{3} \frac{6+16 \lambda_*}{7+16 \lambda_*}$\;  \\ \hline
\end{tabular}
\caption{The coordinates of the four fixed points appearing in the bulk-boundary system extending the Einstein-Hilbert truncation. The values of $g_*, \lambda_*$ are given in \eqref{UVG}.
\label{b.fix}}
\end{center}
\end{table}
The last two fixed points are characterized by the critical exponents
\be
\begin{array}{llll}
\mbox{NG-G-FP:} \qquad & \theta_{1,2} = 1.475 \pm 3.043 i \, , \qquad & \theta_3 = -2 \, , \qquad & \theta_4 = 1 \,  \\[1.2ex]
\mbox{NG-NG-FP:} \qquad & \theta_{1,2} = 1.475 \pm 3.043 i \, , \qquad & \theta_3 = 3 \, , \qquad  & \theta_4 = 2 \, .
\end{array} 
\ee
Thus, at the NG-NG-FP all four scaling fields are UV relevant while the NG-G-FP exhibits one irrelevant scaling field.

\subsubsection*{The semi-classical approximation}
In view of phenomenological applications it is useful to study the full system in the semi-classical (or one-loop)
approximation. In this case, the analysis neglects the ``improvement terms'' containing the anomalous dimensions $\eta_N$ and $\eta_N^\p$ 
on the r.h.s.\ of the flow equation \eqref{FRGE} and all threshold functions are evaluated at zero cosmological constant.
In this manner the approximate anomalous dimensions are given by
\be
\eta_N = - (d-2) \, \omega_d \, g \, , \qquad \eta_N = - (d-2) \, \omega_d^\p \, g^\p \, ,
\ee
with the numerical coefficients
\be
\begin{split}
\omega_d = & \, \frac{1}{3(d-2)(4\pi)^{d/2-1}} \left( ( 6d(d-1) +24) \Phi^2_{d/2}(0) - d(d-3) \Phi^1_{d/2-1}(0) \right) \, , \\
\omega_d^\p = & \, - \frac{d(d-3)}{3(d-2)(4\pi)^{d/2-1}} \, \Phi^1_{d/2-1}(0)  \, . 
\end{split}
\ee
For $d=4$ and the optimized threshold functions these evaluate to
\be\label{omegaev}
\omega_4 = \frac{11}{6\pi} > 0 \, , \qquad \omega_4^\p = - \frac{1}{6\pi} < 0 \, . 
\ee

The dimensionful RG equations $k \p_k G_k = \eta_N \, G_k$ then have the following simple but exact solution
\bea\label{Gbulkk}
G_k & = & \frac{G_0}{1+ \omega_d \, G_0 \, k^{d-2}}\,, \\ \label{Gboundk}
G_k^\p & = & \frac{G_0^\p}{1+ \omega_d^\p \, G_0^\p \, k^{d-2}}\,.
\eea
The signs in \eqref{omegaev} imply hat the bulk and boundary Newton's constants run in opposite 
directions. Since $\omega_4$ is positive, $G_k$ \emph{decreases} for increasing $k$ while
$\omega_4^\p<0$ implies that $G_k^\p$ \emph{increases} when $k$ is increased. The feature
that $\eta_N$ and $\eta_N^\p$ come with different signs also holds beyond the semi-classical approximation
and is largely independent of the choice made for $\cR_k$. The relation \eqref{Gbulkk} interpolates
continuously between $\lim_{k \rightarrow 0} G_k = G_0$ and $\lim_{k \rightarrow \infty} g_k = (\omega_d)^{-1}$,
respectively.

When studying the consequences arising from the phase diagram 
 shown in figure \ref{EHflow}, it is important to realize that while many features
 like the existence and critical exponents of the fixed points and the crossover pattern
 is independent of the choice of regulator, the detailed form of the RG trajectory
will depend on $\cR_k$. From the phenomenological point of view it is then 
viable to approximate the RG trajectories through analytic formulas capturing
the universal features of the flow. 
Restricting to $d=4$, one parameterization, initially advocated in Refs.\ \cite{Koch:2010nn,Koch:2013rwa} models
the RG flow of $G_k$ and $\Lambda_k$ by the two-parameter family of curves
\be\label{lvong}
g(k)= \frac{G_0 k^2}{1+k^2/g_*} \quad,
\ee
\bea\label{lambdalit}
\lambda(k)&=&
\lambda_{*}+\frac{1}{k^2} \Lambda_0
-\frac{g_* \lambda_*}{G_0 k^2}
\text{Log}\left[ \left(1+G_0\frac{k^2}{g_*}\right)\right]\quad.
\eea
These interpolate smoothly between $\lim_{k \rightarrow \infty} (g_k, \lambda_k) = (g_*, \lambda_*)$
in the UV and the two free parameters $\lim_{k \rightarrow 0} (G_k, \Lambda_k) = (G_0, \Lambda_0)$
in the IR, thus reproducing the correct crossover behavior from the quantum to the classical regime.
 We will mostly resort to either \eqref{Gbulkk} or this parameterization of the RG trajectories
when performing numerical studies with $\lambda_k$ and $g_k$ in the sequel.

Based on the flow pattern shown in figure \ref{EHflow}, it is plausible that the short distance physics of a black hole is not controlled
by classical general relativity but intrinsically connected to the NGFP, entailing the scaling relations $\eqref{UVG}$. Notably, these
scaling relations are independent of the Einstein-Hilbert truncation used to derive the beta functions. 
They solely rely on the existence of the NGFP and the resulting scale invariance at this point. We will then
review consequences resulting from the scale dependent coupling constants for the structure and singularities of the classical black hole solution in the following sections.  

\subsection{The physical interpretation of the RG scale}
\label{secScaleSet}
The scale dependence of the coupling constants investigated
in the last section immediately carries over into
a scale dependence of the effective average action.
A key question when dealing with a concrete physics problem then concerns
the relation of the cutoff scale $k$ to the physical scales
of the system under consideration.
For example, the dimensionful physical constants for a classical Schwarzschild
black hole are Newton's coupling $G_0$ and the black hole mass $M$.
Thus, when calculating quantum corrections to this system by using $\Gamma_k$
 one expects that the scale $k$ is to be identified with a ``typical''
 scale of the geometry and in addition depends parametrically on the
dimensionful parameters of the system $k=k(x; G_0,M_0)$.

In general, the cutoff identification $k = k(x)$ is not unique and its construction always involves a certain level of ``physics intuition''. There are, however, some very general principles that any physically reasonable choice should satisfy. Firstly, the function $k(x)$ should be independent of the particular choice of coordinates in which the (black hole) geometry is formulated. In other words, $k(x)$ should be constructed from diffeomorphism invariant quantities as, e.g., proper distances or scalar curvature invariants. Secondly, it is reasonable to expect that short distance physics should map to large values $k$ while the long distance physics should be related to small values of $k$. Finally, the cutoff identification should respect the symmetries of the underlying geometry, i.e., a Killing vector of the classical geometry should remain a Killing vector upon carrying out the improvement procedure. For problems that involve more than one dimensionful scale the implementation of these requirements still leaves some freedom in the construction of $k(x)$, 
so that the robustness of the conclusions drawn from the RG improvement has to be checked for various plausible choices of the cutoff identification. 
In the following we will then introduce the most frequently used scale setting procedures in asymptotic safety,
where only the latter ones have been applied in the context of black hole geometries.

A general scale setting procedure 
interprets the $k$ in $\Gamma_k$ as an independent field and
subsequently solves the corresponding 
 (algebraic) equations of motion \cite{Reuter:2003ca,Reuter:2005kb,Koch:2010nn,Domazet:2012tw}.
For the the Einstein-Hilbert truncation this procedure
corresponds to solving the equation
\be\label{consistency}
R\left( \frac{1}{G_k}\right)'-2 \Lambda_k \left( \frac{\Lambda_k}{G_k}\right)'=0 \quad,
\ee
where $'\equiv \partial_{k^2}$.
Close to the UV fixed point one approximates $G_k\approx g_{*} /k^2$ and 
$\Lambda_k \approx \lambda_{*} k^2$ and equation (\ref{consistency}) gives
\be\label{UVconsistency}
k^2= \frac{R}{4 \lambda_*} \quad.
\ee

An other general procedure is motivated on the observation that 
$k^2$ can in principle be identified with any diffeomorphism-invariant quantity that is present in
the theory, that has the right dimensions. It was argued in Refs.\ 
\cite{Frolov:2011ys,Bonanno:2012jy,Hindmarsh:2012rc,Copeland:2013vva}
that the simplest choice along this line
would be
\be\label{Rchoice}
k^2=\xi R \quad,
\ee
which is actually consistent with the UV limit of the condition (\ref{consistency}).

Ideally, when applying this procedure, the equations of motion of the improved action, 
must be solved simultaneously with the scale setting condition.
For example in the Einstein Hilbert case without matter, the equations
\begin{eqnarray}
 R_{\mu\nu} - \frac{1}{2} g_{\mu\nu} R=-g_{\mu\nu}\Lambda_{k}-
G_{k}\left(g_{\mu\nu} \, D^2-D_\mu D_\nu\right)\frac{1}{G_{k}}\quad , 
\label{eomImproved}
\end{eqnarray}
would have to be solved simultaneously for $g_{\mu \nu}(x)$ and $k(x)$,
since the condition (\ref{consistency}) is actually the Bianchi identity
applied to the equation (\ref{eomImproved}).
Due to the highly non-trivial form of the functional dependence of $\Lambda_k$ and $G_k$,
this procedure has however not yet been realized for black holes
where, so far, an alternative
improvement method has been used.

An alternative tool of studying effects of running couplings in quantum field theory
is the procedure of improving classical solutions by 
``upgrading'' the coupling constants to their scale-dependent
counterparts. The parameter $k$ introduced in this manner 
must then be
connected to the physical quantities of the system through a scale setting procedure. 
This idea has for example been successfully implemented in the context of QED
when deriving the Uehling correction to the Coulomb potential \cite{Dittrich}.
For the case of gravity one 
could try to use one of the scale setting prescriptions (\ref{consistency}, \ref{Rchoice}), however
their application to specific physical problems and improvement procedures is not ideal.
For the case of improving classical black hole solutions this can be easily 
seen by the fact that the classical Schwarzschild solution actually has a vanishing Ricci
scalar everywhere outside of the black hole $R=0$.
Thus applying (\ref{Rchoice}) would actually not improve the classical solution at all
since $k$ would simply remain zero. 

In order to obtain a finite and non-zero scale identification for 
improving classical black hole solutions one typically refers
to a different scale setting scheme. This 
 procedure  assumes that for dimensional
reasons the momentum scale $k$ has to be inversely proportional to a physical distance
scale $d(r)$ of the classical geometry
\be\label{kvonP}
k(P(r))=\frac{\xi}{d(P(r))}\quad,
\ee
where $d(P)$ is the distance scale which provides the relevant cutoff $k$ when 
the test particle is located at the point $P$. The constant
$\xi$ is expected to be of order unity and has to be fixed by an additional 
physics argument. Still, there are various ways of defining a meaningful physical
distance scale $d(P(r))$ for black holes.
In the context of RG-improved Schwarzschild black holes, several choices for $d(r)$ have
been advocated in the literature. Here we will identify $d(P)$ with the radial proper distance
from the center of the black hole to the point $P$ along a
purely radial curve ${\mathcal{C}}_r$ \cite{Bonanno:2000ep}
\be\label{dproperr}
d_r(r)=\int_{{\mathcal{C}}_r} \sqrt{|ds^2|} \, .
\ee
This prescription has the nice feature that it allows for a very direct physical interpretation
in terms of distance outside the horizon. A possible caveat comes due to the
fact that $ds$ actually changes its sign at the horizon. A monotonic $d(r)$ is recovered by imposing the
condition of an absolute value $|ds|$ in the integral.
An alternative, equally reasonable, cutoff identification uses
the proper time it takes for an observer starting at rest
at coordinate distance $r$ from the black hole singularity to reach $r = 0$
\be\label{dpropert}
d_t(r)=\int_{{\mathcal{C}}_t} \sqrt{ds^2} \, .
\ee
This proper time identification building on the radial timelike
geodesic ${\mathcal{C}}_t$ has the advantage that it is  smooth
at the classical horizon.

\section{RG-improved black holes in asymptotically flat space}
\label{sect.3}
After the short review of the fundamental ingredients underlying asymptotic safety,
we now apply the RG-improvement scheme introduced in section \ref{secScaleSet}
to the classical black hole solutions in order to get some first insights
in the quantum properties of black holes within asymptotic safety.
We start with the improved Schwarzschild solution in section \ref{subsecSSS}
and subsequently generalize the discussion by including 
angular momentum (section \ref{sect:3.2}) and extra dimensions (section \ref{sect:3.3}). 
\subsection{The Schwarzschild solution}
\label{subsecSSS}

The classical Schwarzschild black hole is described
by the metric \eqref{ssmetric} with the radial function (\ref{frfct}).
As described in section \ref{secScaleSet} this classical solution
of the Einstein equations can be quantum-improved,
by replacing the classical coupling constants
by their scale-dependent counterparts.
Thus, the improved radial function reads
\be\label{frimpf}
f(r)=1- \frac{2 \, G_k \, M}{r}\quad.
\ee
The precise details of the scale setting $k_i(r) = \xi/d_i(r)$ then depend on the choice of the
spacetime curve that is used for the definition of the physical cutoff.
For the purely radial curve ${\mathcal{C}}_r$ the integration gives inside the 
classical horizon
\be\label{d1in}
d_{r}(r)|_{r<2G_0 M}= 2 G_0 M \arctan \sqrt{\frac{r}{2 G_0 M-r}}-\sqrt{r(2 G_0 M-r)}
\ee
and outside the classical horizon
\be\label{d1out}
d_{r}(r)|_{r>2G_0 M}=
\pi G_0 M + 2 G_0 M \ln \left( \sqrt{\frac{r}{2 G_0 M}}\sqrt{\frac{r}{2 G_0 M}-1}\right)
+\sqrt{r(r-2 G_0 M)}\quad.
\ee
Eqs.\ (\ref{d1in}) and (\ref{d1out}) match at the classical horizon.
For a more compact description it is useful to extract the asymptotic behavior
of these distance functions. Close to the curvature singularity at $r=0$
one finds
\be\label{d1inin}
d_{r}(r)|_{r\ll 2G_0 M} \simeq \frac{2}{3}\frac{1}{\sqrt{2 G_0 M}} r^{3/2} + {\mathcal{O}}(r^{5/2})
\ee
while the asymptotic behavior for large $r$ is given by
\be\label{d2outout}
d_{r}(r)|_{r\gg 2G_0 M} \simeq r+ {\mathcal{O}}(r^{0}) \, . 
\ee
The distance function obtained from the proper time of a radially infalling observer ${\mathcal{C}}_r$ 
leads to the small $r$-asymptotics 
\be\label{ptimp}
d_{t}(r) \simeq \frac{\pi}{2}\frac{1}{\sqrt{2 G_0 M}} r^{3/2} + \ldots \, .  \quad
\ee
This asymptotic coincides with (\ref{d1inin}) for small $r$ up to an irrelevant numerical factor. Substituting these 
asymptotics into \eqref{kvonP} one obtains
\be\label{krasym}
k(r) \simeq \frac{3}{2} \, \sqrt{2 G_0 M} \,  \xi  \, r^{-3/2} + \ldots \, . 
\ee
The free parameter $\xi$ is typically of order one and the
different normalizations \eqref{d1inin} and \eqref{ptimp} can 
be absorbed by a redefinition of $\xi$. Thus the radial and proper time improvements
coincide in the UV (small radii), but the latter leads
to a lower scale choice $k_t(r)<k_r(r)$ in the IR (large radii).
When performing a RG improvement for a particle physics system
which admits more than one intrinsic momentum scale, one chooses typically the largest scale that can cut off the running couplings.
Using this analogy, the most commonly used scale identification
employs ${\mathcal{C}}_r$. While the cutoff identification originating from
the distance functions (\ref{d1in}, \ref{d1out}) is perfectly
valid, it is convenient to approximate the distance function
by 
\be\label{interpol}
d_r(r)=\sqrt \frac{2 r^3}{2r+9 G_0 M }\,.
\ee
This rather simple analytic formula interpolates smoothly between the two scaling regimes obtained as $r\rightarrow 0$
and $r \rightarrow \infty$. 

We now insert the cutoff-identification $k(r) = \xi/d_r(r)$
into the improved solution \eqref{frimpf}. Here it would be straightforward
to carry out this improvement based on the scale-dependence 
encoded in the full beta functions \eqref{betaeq} using
numerical methods. In order to get analytic access to 
the system, one can, however, approximate
the scale-dependence of Newton's constant by the one-loop
result \eqref{Gbulkk}. Combining this flow with the distance function
\eqref{interpol} 
one obtains the improved radial function
\be\label{frSSImp}
f(r)=1-\frac{4 G_0 M r^2}{2 r^3 + \tilde \omega G_0(2r+9G_0 M)}\quad,
\ee
where $\tilde \omega=\omega \xi^2$.
The improved radial function determines the properties
of the quantum-improved black hole such as its horizon structure, asymptotic behavior and thermodynamics.

\begin{figure}[t!]
\centerline{\includegraphics[width=12cm]{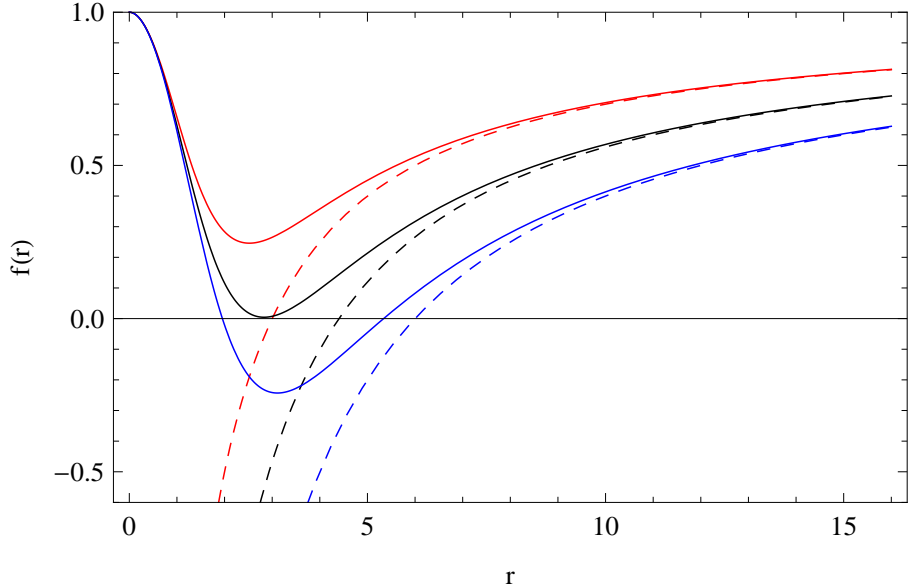}}
\caption{\label{figfvonrSS} Radial dependence of the laps function $f(r)$ for $G_0=1,\, \tilde \omega=1$ for different values
of the mass $M=1.5, \,2.2,\, 3$ (red top, black middle, blue bottom). The classical behavior represented by the dashed lines is 
shown for comparison.}
\end{figure}

\begin{figure}[t!]
\centerline{\includegraphics[width=4cm]{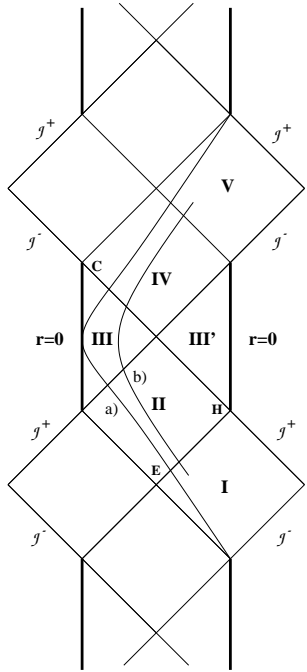}}
\caption{\label{figPenroseSS}
Penrose diagram for the improved Schwarzschild spacetime 
\cite{Bonanno:2000ep}. Depicted are two geodesics of radially infalling particles, 
one of a particle with $K_{kin}>0$ at $r\rightarrow \infty$ and an other one
of a particle starting at rest at finite $r$. One observes that the latter does not reach the origin at $r=0$.}
\end{figure}

\begin{figure}[t!]
\centerline{\includegraphics[width=12cm]{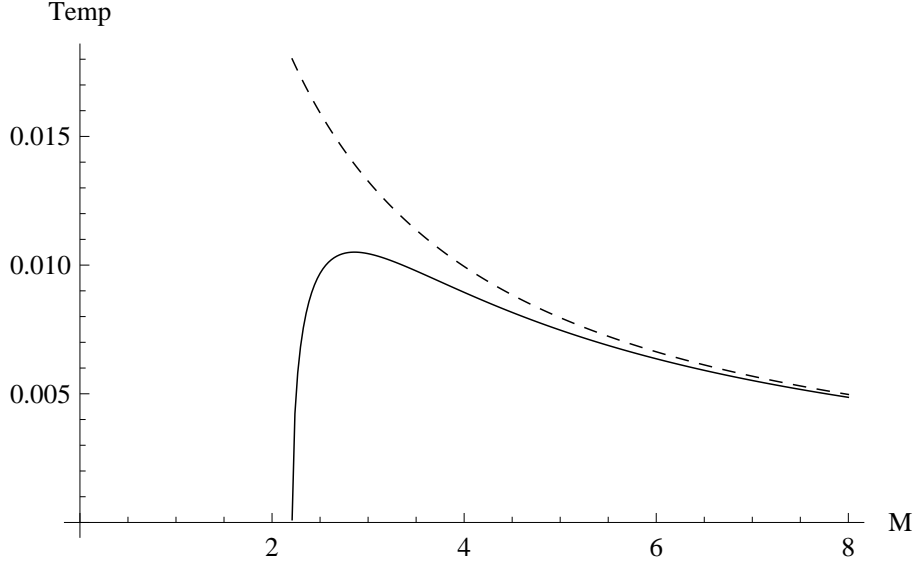}}
\caption{\label{figTvonMSS} Mass dependence of the Hawking temperature
for the improved Schwarzschild solution (solid line) for $G_0=1$ and $\tilde \omega =1$ in Planck units
in comparison to the classical temperature (dashed line).}
\end{figure}

The properties of the quantum-improved black hole resulting from (\ref{frSSImp}) 
are conveniently obtained from figure \ref{figfvonrSS}
which shows the improved radial function for various masses $M$.
This analysis leads to the following conclusions: \\[1.2ex]
{\bf Classical limit:} \\
First of all one sees that for large $r$ the improved line-element agrees with the classical one
\be\label{frSSImpApr}
f(r)= 1- \frac{2 G_0 M}{r}\left( 1-\tilde \omega \frac{G_0}{r^2}\right) +{\mathcal{O}}\left(\frac{1}{r^4}\right)\, .
\ee
This can also be stipulated from the large $r$ behavior of figure \ref{figfvonrSS}.
Furthermore, (\ref{frSSImpApr}) allows to read off the leading correction to Newton's potential.
For $\tilde \omega=118/15 \pi$ this correction is in agreement
with the perturbative result obtained in Ref.\ \cite{Donoghue:1993eb}. This 
agreement can then be used to fix the (hitherto undetermined) parameter $\xi$. \\[1.2ex]
{\bf Singularity:}\\
The classical Schwarzschild black hole possesses a curvature singularity at $r=0$
where the square of the Riemann tensor diverges. 
For the improved solution the fate of this singularity can be investigated
by expanding (\ref{frSSImp}) for small $r$
\be\label{frSSImpApr2}
f(r) \simeq 1-\frac{4 r^2}{9 \tilde \omega} +{\mathcal{O}}(r^3)\quad.
\ee
This shows that the improved radial function is actually well-behaved
at $r=0$ and has the form of the de Sitter metric with an effective
cosmological constant $\Lambda_{\rm eff} = 4/(3 \tilde \omega)>0$.
This feature was baptized ``de Sitter core'' in Ref.\ \cite{Bonanno:2000ep}.
The regularity of the improved solution can furthermore be verified
by calculating the invariants
that are singular for the classical Schwarzschild black hole
\be
R_{\mu \nu \alpha \beta}R^{\mu \nu \alpha \beta}=2 \left( \frac{4}{3}\right)^3\tilde \omega^2\quad.
\ee
Thus the square of the Riemann tensor remains finite, pointing at the resolution
of the classical black hole singularity. \\[1.2ex]
{\bf Horizon structure}:\\
The horizon structure encoded in the improved radial function (\ref{frSSImp})
turns out to be richer than in the classical case. One finds
that the number of horizons $N_H$ depends on the mass parameter
\be
N_H=\left\{
\begin{array}{ccc}
0& {\mbox{for}}& M< M_{cr}\\
1 ({\mbox{double}})& {\mbox{for}}& M= M_{cr}\\
2& {\mbox{for}}& M>M_{cr}\\
\end{array}
\right. \quad,
\ee
as it can be seen from figure \ref{figfvonrSS}. For $M > M_{cr}$ there is an outer and an inner horizon situated at $r_+$ and $r_-$, respectively. If the mass of the black 
hole equals the critical mass, the two horizons coincide while for $M < M_{cr}$ there is a naked singularity.
For the interpolation (\ref{interpol}), the critical mass $M_{cr}$
is found to be $M_{cr}\approx 0.20$ in Planck units. \\[1.2ex]
{\bf Causal Structure}: \\
Apart from the non-singular structure at $r=0$, the causal structure implied by the improved radial function 
is very similar to the one of a Reissner-Nordstr\"om black hole.
For $M>M_{cr}$, one can distinguish five main regions
\bea
I \; {\mbox{and}}\; V: & r_+<r<\infty \\ \nonumber
II \; {\mbox{and}}\; IV: & r_-<r<r_+ \\ \nonumber
III \; {\mbox{and}}\; III': & 0<r<r_- \quad.
\eea
It is straight forward to calculate the geodesics in this spacetime structure.
A typical geodesic is shown in the Penrose diagram, figure \ref{figPenroseSS}. 
Starting in region I ($r<\infty$) at rest
the test mass will fall into the black hole, pass the horizon $r_+$ and transit the region $II$ with inverted time direction.
When passing region III, the test mass is repelled outwards, back to regions with larger $r$: IV and V.
This is in full analogy to the Reissner-Nordstr\"om case. However, one difference is that the improved lapse function
allows a particle with sufficient energy ($K_{kin}>0$ at $r\rightarrow \infty$) to actually reach the non-singular point $r=0$. \\[1.2ex]
{\bf Thermodynamics}: \\
Following the standard discussion of black hole thermodynamics in Euclidean time and
with the usual regularity condition at the cone of the resulting $R^2\times S^2$ topology,
the temperature of a horizon is given by \eqref{HT}. Applying
this equation to  (\ref{frSSImp})
gives the temperature at the outer black hole horizon at $r_+$
\be
T_{BH}=\frac{1}{8 \pi} \frac{M G_0 r_+ (r_+^3-\tilde \omega G_0 r_+-\tilde \omega G_0^2 9 M)}{
(r_+^3+\tilde G_0 (r_++(9/2) G_0 M))^2} \quad.
\ee
Since also the horizon radius $r_+$ is a function of $M$,
it is instructive to plot the temperature $T_{BH}(M)$ shown in figure \ref{figTvonMSS}.
One observes that for masses $M>M_{cr}$ the temperature follows the same
$1/M$ dependence as for the usual classical solution. However,
when $M$ approaches $M_{cr}$, the temperature of the improved solution
drops to zero, which was be interpreted as the formation of a stable black hole remnant.
This picture of remnant formation has later on also been confirmed
by analyzing the dynamics of the black hole evaporation process 
based on the RG improved Vaidya-metric \cite{Bonanno:2006eu}.
%

\subsection{The Kerr solution}
\label{sect:3.2}
We now review the extension of the previous analysis to black holes
carrying angular momentum, thereby mainly following the original Refs.\ \cite{Reuter:2006rg,Reuter:2010xb}.
The classical rotating black hole in four dimensions is described by
the Kerr metric (reviewed, e.g., in Ref.\ \cite{Taylor}). In Boyer-Lindquist coordinates the classical line element of
this metric is
\begin{equation}
ds^{2}=-\left( 1-\tfrac{2MG_{0}r}{\rho^2 }\right) dt^{2}+\tfrac{%
\rho ^{2}}{\Delta }dr^{2}+\rho ^{2}d\theta ^{2}+\tfrac{\Sigma \sin ^{2}\theta
}{\rho ^{2}}d\varphi ^{2}-\frac{4MG_{0}ra\sin ^{2}\theta }{\rho ^{2}}%
dtd\varphi . \label{dsKerr}
\end{equation}
Here $a = J/M$ is the reduced angular momentum and we use the common abbreviations
\begin{equation} \label{rhoKerr}
\rho ^{2}\equiv r^{2}+a^{2}\cos ^{2}\theta \,, \qquad 
\Delta \equiv r^{2}+a^{2}-2MG_{0}r  \,,  
\end{equation}
and
\begin{equation}
\Sigma \equiv \left( r^{2}+a^{2}\right) ^{2}-a^{2}\Delta \sin ^{2}\theta \, . 
\label{SigmaKerr}
\end{equation}
The family of Kerr solutions are characterized by two parameters the mass $M$ and their angular momentum $J$.
The non-trivial angular momentum reduces the spherical symmetry, characterizing the Schwarzschild solution, to axial symmetry
around the rotation axis. As a consequence the Kerr solution has 
 has a richer structure than the Schwarzschild case. The $g_{rr}$-component degenerates at the two (spherically symmetric) 
  horizons situated at
\begin{equation}
r_{\pm }=M G_0\pm \sqrt{(M G_0)^{2}-a^{2}}  \,.\label{rHKerr}
\end{equation}
The vanishing of $g_{tt}$ produces two static limit surfaces $S_{\pm }$ at
\begin{equation}
r_{S_{\pm }}\left( \theta \right) =M G_0\pm \sqrt{(M G_0)^{2}-a^{2}\cos ^{2}\theta }\,.
\label{rSKerr}
\end{equation}
Owed to the angular momentum, the horizons and limit surfaces coincide at the poles $\theta = 0, \pi$ only.
Moreover, the classical Kerr solution possesses two translational Killing vectors related to
the time direction and one angular direction
\begin{equation}
\mathbf{t}\equiv t^{\mu }\partial _{\mu }=\frac{\partial }{\partial t}\, ,\qquad %
\boldsymbol{\varphi}\equiv \varphi ^{\mu }\partial _{\mu }=\frac{\partial }{%
\partial \varphi }  \, . \label{KillingKerr}
\end{equation}

\begin{figure}[t!]
\centerline{\includegraphics[width=10cm]{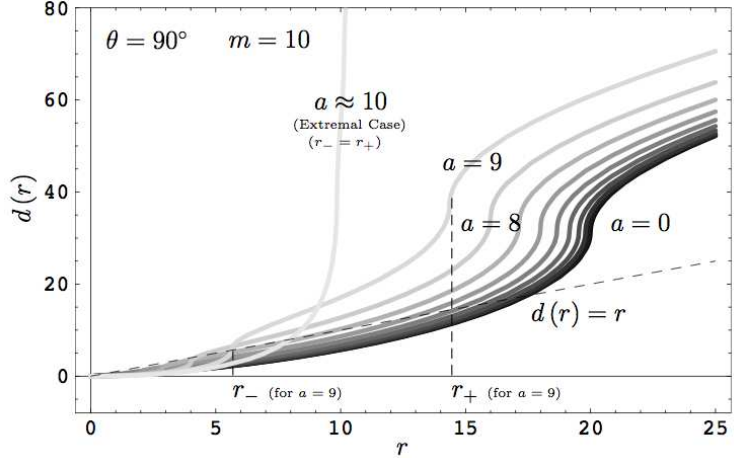}}
\caption{\label{figdrKerr} Distance scale $d(r)$ in Planck units 
at $\theta=\pi/2$, for $M G_0=10$, and for various values of $a$
 \cite{Reuter:2010xb}. For comparison this figure further shows
 the straight line that would correspond to $d(r)=r$.}
\end{figure}

In the context of asymptotic safety, this classical Kerr spacetime can be improved
in the same spirit as the Schwarzschild black hole. Still maintaining the general guideline
that the RG improvement should not break the classical symmetries of the solution, 
the reduced symmetry group of the Kerr black hole allows an additional $\theta$-dependence 
in the cutoff identification
\be\label{kvonrt}
k=k(r,\theta)=\xi/d(r,\theta) \, .
\ee
One choice for the distance function  then
generalizes the scale setting procedure (\ref{kvonP})
to the classical Kerr metric \eqref{dsKerr} by taking the (absolute value)
of the proper distance between the origin and the point $P$ as $d(r,\theta)$.
The natural choice for the curve $\cC_r$ connecting these points is the ``shortest path'' with $dt=d\phi=d\theta=0$ 
leading to the identification
\begin{equation}
d\left( r,\theta \right) =\int_{0}^{r}d\bar{r}\sqrt{\left| \frac{\bar{r}%
^{2}+a^{2}\cos ^{2}\theta }{\bar{r}^{2}+a^{2}-2 \, G_0 \, M \, \bar{r}}\right| }\quad.
\label{drKerr}
\end{equation}
The distance function is easily obtained by a numerical evaluation of the integral.
For special cases as, e.g., in the equatorial plane analytical expressions can be
obtained as well. The distance function obtained in this way is illustrated in figure \ref{figdrKerr}, which shows the $r$-dependence of $d(r,\theta)$ 
 in the equatorial plane.
One readily observes that at the two horizons $r_\pm$ the first derivative of $d(r,\theta)$
diverges while the function itself remains finite. Moreover, the two horizons
approach each other for increasing $a$. For the critical value
$a = M G_0$, the two horizons coincide, 
leading to a divergence in the distance function at $r_\pm$. 
For $\theta \neq \pi/2$ the $d(r,\theta)$ obtained numerically
 displays the same qualitative behavior albeit with larger
numerical values $d(r,\theta = \pi/2) < d(r,\theta \not = \pi/2)$. 
Since the $\theta$ dependence turns out to be rather weak,
the following discussions will concentrate on the $r$ dependence of the energy scale $k$.

The explicit form of the RG-improved
Kerr metric is obtained by replacing $G_0\rightarrow G_{k}$
in the classical line element. Again we use the one-loop approximation
\eqref{Gbulkk} to parameterize the scale-dependence of Newton's constant.
The cutoff identification \eqref{kvonrt} supplemented by the distance function
\eqref{drKerr} then allows to translate this $k$-dependence
into a dependence on $r$ (and $\theta$) leading 
to the RG-improved
Kerr metric 
\begin{equation}
ds^{2}=-\left( 1-\tfrac{2MG_{k(r)}r}{\rho^2 }\right) dt^{2}+\tfrac{%
\rho ^{2}}{\Delta_I }dr^{2}+\rho ^{2}d\theta ^{2}+\tfrac{\Sigma_I \sin^{2}\theta
}{\rho ^{2}}d\varphi ^{2}-\tfrac{4MG_{k(r)}ra\sin ^{2}\theta }{\rho ^{2}}%
dtd\varphi . \label{dsKerrI}
\end{equation}
Here, $\Delta_I$ and $\Sigma_I$ denote the improved counterparts of the functions \eqref{rhoKerr}
and \eqref{SigmaKerr}, respectively
\begin{equation}\label{DeltaKerrI}
\Delta_I \equiv r^{2}+a^{2}-2MG_{k(r)}r  \,, \qquad
\Sigma_I \equiv \left( r^{2}+a^{2}\right) ^{2}-a^{2}\Delta_I \sin ^{2}\theta \, . 
\end{equation}

\begin{figure}[t!]
\centerline{\includegraphics[width=10cm]{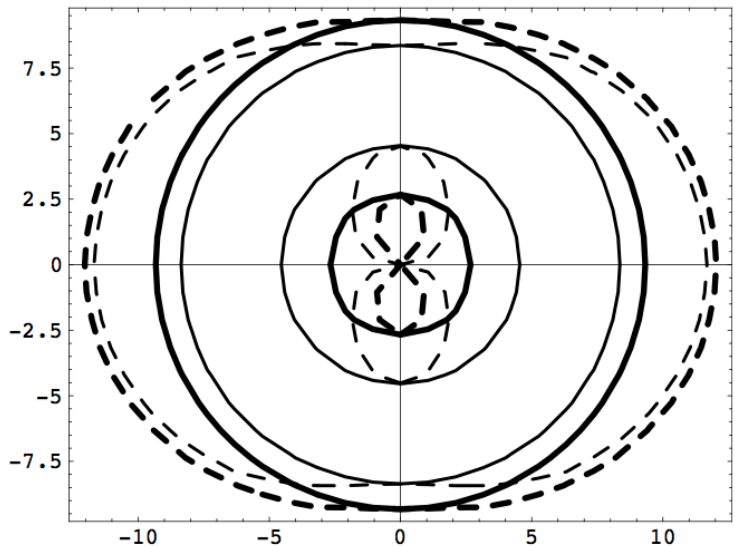}}
\caption{\label{figHorizonsKerr} Structure of a classical Kerr black hole (thick lines) with $G_0 M=6 $ and
$a=5$ (in Planck units), 
in the $x$-$y$--plane, in comparison to its improved counterpart (non-thick lines).
 Solid lines represent event horizons, while dashed lines represent the static limit surfaces \cite{Reuter:2010xb}.}
\end{figure}

 Several features of the RG-improved Kerr spacetime can be deduced by studying special classes of observers.
The conserved quantities for trajectories in the improved Kerr solution can be obtained
from $u^\mu \equiv dx^\mu/d\tau$, the momentum $p^\mu=mu^\mu$ and the Killing
vectors (\ref{KillingKerr})
\be\label{conservedKerr}
E= -t_\mu p^\mu \, , \qquad 
L= -\phi_\mu p^\mu \, . 
\ee
\begin{itemize}
\item
A special case of observers is given by the condition $L=0$, which is known as the
zero angular momentum observer.
Although the angular momentum of this observer is vanishing,
the relation (\ref{conservedKerr}) gives a non-vanishing temporal
dependence of the angular coordinate itself
\be
\omega\equiv \frac{d\phi}{dt}=\frac{2 \, G(r) \, M \, a \, r}{ a^{2}\Delta_I(r) \sin ^{2}\theta-\left( r^{2}+a^{2}\right) ^{2}}\quad.
\ee
For constant $G_0$ and $\Delta$, this effect is known as frame dragging. For 
 the improved Kerr black hole $\omega$  is modified due to the radial dependence
of $G(r)$ and $\Delta_I(r)$. 
\item
The RG improvement also modifies the classical static limit surfaces \eqref{rSKerr}.
The condition $t^\mu t^\nu g_{\mu \nu}=0$ is then equivalent to 
\be\label{condSLKerr}
Q_S(\tilde \omega,\,a,\,M,\,\theta)\equiv r^2+a^2\cos^2\theta-2 \, G(r)\, M\, r=0 \, .
\ee
Solving this relation for $r$ gives the improved static limit surfaces $S^I_{\pm}$
situated at $r_{S\pm}^I =r_{S\pm}^I(\theta)$.
Like in the classical case, the number of static limit surfaces 
depends on the values of $M$ and $a$. One observes that the variable coupling 
$G(r)$ comes with the mass parameter $M$. Therefore, this correction, as long
as it is small, can be captured by
a ``renormalized''  (somewhat lower) effective mass $\tilde M \approx M G(r_\pm)/G_0$.
\item
In order to study the event horizon of (\ref{rSKerr}),
one has to study so called stationary observers, which perceive no time
variation of the gravitational field. This can be implemented by demanding
that the observer's four velocity $u^\mu$ is proportional to the Killing vector $ \xi^\mu= t^\mu+(d\phi/dt) \phi^\mu$,
say $u^\mu=\gamma \xi^\mu$.
For large black hole masses $M\gg m_{Pl}$, this condition leads to
\be\label{condHorKerr}
Q_H(\tilde \omega,\,a,\,M)= r^2+a^2-2 \, G(r) \, M \, r=0 \quad.
\ee
The solutions of this relation determine the position $r_{H \pm}^I$ of the improved horizons. 
\end{itemize}
For the classical Kerr solution, as well as the improved solutions where the improvement can be considered as a
small correction, the four surfaces can be ordered by increasing radius
\be
r_{S-}(\theta)\le r_{-}\le r_{+}\le r_{S+}(\theta) \,.
\ee
This rich structure of horizons and static limit surfaces is illustrated in figure \ref{figHorizonsKerr}.
One observes that due to the improvement, the outer horizon and static limit surface ($r_+,\,r_{S+}$) gets
shifted inward, while the inner horizon and static limit surface ($r_-,\,r_{S-}$) are pushed towards larger values $r$.

Classical Kerr black holes with $a > G_0 \, M$ have no horizons and give rise to a naked singularity. 
In order to prevent this problem one imposes a so called extremality condition on those parameters.
This extremality occurs when the horizons $H_\pm$ coincide at a radius $r_e$. For the improved black
hole, this extremality condition can be based on \eqref{condHorKerr} and translates to the conditions 
\bea \label{criticalityKerr}
Q_H(\tilde \omega,\,a,\,M)|_{r_e}&=&0 \,, \\ \nonumber
\frac{d Q_H(\tilde \omega,\,a,\,M)}{d r}|_{r_e}&=&0 \,, \\ \nonumber
\frac{d^2 Q_H(\tilde \omega,\,a,\,M)}{d r^2}|_{r_e}&\ge &0 \,.
\eea
For the classical case, the solution of the conditions (\ref{criticalityKerr}) can 
be found analytically and gives a linear dependence between the mass $M$ and the angular
momentum parameter $a$: the critical black hole satisfies $a = G_0 M$.
The RG-improved solution shows this
behavior for large values $a$ only. For small $a$ there are sizable
corrections and the critical configuration 
is obtained for finite mass $M_{cr}$. Notably, this result is in 
complete agreement
with the remnant formation picture developed from the analysis
of the Schwarzschild black hole of the previous section
where one also finds a critical configuration at finite remnant mass $M_{cr}$.

An other interesting property of Kerr black holes is the possibility of extracting
energy from them by means of the Penrose process \cite{Taylor}.
This process takes advantage of the fact that the energy of a particle (\ref{conservedKerr})
can actually take negative values. By letting a combined system $A$ fall into the
ergosphere of a Kerr black hole and then letting it disintegrate into two particles
$B$ and $C$, where $B$ falls into the black hole (with negative energy)
and $C$ escapes into infinity, one can show that due to energy conservation
of the entire system, the black hole has actually lost energy.
For large black hole masses $M\gg m_{Pl}$, the discussion of this process follows
the classical discussion with negligible corrections.
The numerical analysis of Ref.\ \cite{Reuter:2010xb} shows that quantum effects only turn out to become important for relatively
small black holes masses $M \approx  m_{Pl}$.
The suppression of this kind of energy loss is analogous to the suppression of the
Hawking radiation in the Schwarzschild case, as it was discussed in section \ref{subsecSSS}.

As it was previously mentioned, many corrections due to renormalization group
effects can be interpreted in terms of renormalized classical ``charges''.
For the case of the Kerr solution, the classical charges are the mass parameter $M$
and the angular momentum $J$.
These charges can also be obtained from surface integrals, by calculating the so-called Komar integrals
\begin{equation}
M_{\text{Komar}}=-\frac{1}{8\pi G_{0}}\int_{S}\nabla ^{\alpha }t^{\beta
}dS_{\alpha \beta } \quad, \label{MKomar}
\end{equation}
and
\begin{equation}
J_{\text{Komar}}=\frac{1}{16\pi G_{0}}\int_{S}\nabla ^{\alpha }\varphi
^{\beta }dS_{\alpha \beta }  \quad,\label{JKomar}
\end{equation}
for the given metric.  Here $S$ is a two-sphere at spatial infinity. The surface element $%
dS_{\alpha \beta }$ is given by $dS_{\alpha \beta }=-2n_{\left[ \alpha
\right. }r_{\left. \beta \right] }\sqrt{\sigma }d^{2}\theta $ where $%
n_{\alpha }$ and $r_{\alpha }$ are the timelike and spacelike normals to $S$.
For the classical Kerr solution, the integrals give exactly the corresponding
parameters in the metric \cite{Poisson}. Further, since the improved line element of the
Kerr metric actually approaches the classical line element, on knows that also for the
improved Kerr metric one finds
\begin{equation}
M^I_{\mathrm{Komar}}|_{r\rightarrow \infty}=M\, ,\qquad J^I_{\mathrm{Komar}}|_{r\rightarrow \infty}=J  \quad.
\end{equation}
In the light of this identity, one can deduce that the bare charges ($M,\, J$) and the quantum-improved charges ($M^I_{\mathrm{Komar}},\, J^I_{\mathrm{Komar}}$)
do actually agree at spatial infinity.
The difference with the purely classical case comes due to the observation that
for the classical black hole, the Komar integrals, evaluated at the outer horizon, have 
the same value as at spatial infinity 
\begin{equation}
M_H \equiv M_{\mathrm{Komar}}|_{r\rightarrow r_+}=M\, ,\qquad  J_H\equiv J_{\mathrm{Komar}}|_{r\rightarrow r_+}=J  \label{limMKomarCl}\, .
\end{equation}
This is however not the case for the improved Kerr solution, where the
Komar integrals evaluated at the outer horizon give\cite{Reuter:2010xb}
\begin{equation}
\begin{split}
M_H^I = & M\frac{G\left( r_{+}\right) }{G_{0}}\left\{ 1-\left[ \frac{\left(
r_{+}^{2}+a^{2}\right) G^{\prime }\left( r_{+}\right) }{aG\left(
r_{+}\right) }\right] \arctan \left( \frac{a}{r_{+}}\right) \right\} \\
\label{limMKomar}
J^I_H = & \left\{ J+\left[ 1-\frac{2MG\left( r_{+}\right) }{a}\arctan \left(
\frac{a}{r_{+}}\right) \right] \left[ \frac{M^{2}G^{\prime }\left(
r_{+}\right) r_{+}^{2}}{a}\right] \right\} \frac{G\left( r_{+}\right) }{G_{0}%
}\,. 
\end{split}
\end{equation}
The difference between eqs.\ (\ref{limMKomarCl}) and (\ref{limMKomar}) can readily interpreted
as quantum dressing of the bare charges.
The dressing has various interesting features:\cite{Reuter:2006rg,Reuter:2010xb}
\begin{itemize}
\item
Firstly, one notes that
\be
\frac{M_H^I}{M}\le 1, \; {\mbox{and}} \quad \frac{J_H^I}{J}\le 1 \quad.
\ee
Which means that the ``charge'' at the horizon and the quantum contribution
outside the horizon sum up in such a way that they return to the bare
value at spatial infinity.
\item
Secondly, the relations (\ref{limMKomar}) satisfy the classical Smarr formula\cite{Larry}
\begin{equation}
M_{H}^I=2 \, \Omega _{H} \, J^I_{H}+\frac{\kappa \, A}{4\pi
G_{0}}  \quad, \label{Smarr}
\end{equation}
where 
\be
\Omega_H= a/((r_+^I)^2+a^2) \quad,
\ee
is the angular velocity, 
\be
\kappa=(r_+^I-G(r_+^I) M- r_+^I G'(r_+^I)M)/((r_+^I)^2+a^2)\quad,
\ee
is the surface gravity and 
\be
A = 4 \pi (r_+^2+a^2) \quad,
\ee
 is the surface area
of the black holes outer horizon.
\item
Thirdly, the expressions (\ref{limMKomar}) are formally
identical to the Komar integrals for a classical Kerr-Newman spacetime
with effective electric charge square
\be
q^2=2Mr_{+}^{2}G^{\prime }\left( r_{+}\right) /G_{0} \quad.
\ee
A discussion of the black hole temperature from (\ref{Smarr})
is direct and largely analogous to the discussion in the Schwarzschild case.
\end{itemize}

Although, the horizon and temperature results based on (\ref{Smarr}) turned out
to be straight forward and direct, the formulation of a complete
entropy relation and a first law of black hole thermodynamics seems
to be much harder. Actually it was noted that
``\textit{either} \textit{there
exists no entropy-like state function for the improved (Kerr) black holes or the
classical relation }$T=\kappa /2\pi $ \textit{does not hold true for them}''\cite{Reuter:2010xb}.

\subsection{The Schwarzschild solution with extra dimensions}
\label{sect:3.3}
The discussions of the RG improved Schwarzschild black hole from subsection \ref{subsecSSS} can also be generalized 
to a $d$-dimensional spacetime where $d \ge 4$.
This analysis is of special interest for models with large extra
dimensions \cite{ArkaniHamed:1998rs,Randall:1999vf} where the Planck
scale $M_d$ could be of the order of a few TeV. In this case
energy particle collisions (e.g.\ at the LHC) could lead
to the production of microscopic black holes 
\cite{Banks:1999gd,Giddings:2000ay,Giddings:2001bu,Dimopoulos:2001hw,Hossenfelder:2001dn,Bleicher:2001kh}.
This reduction of the usual Planck scale to
the microscopic Planck scale $M_{d}$ is achieved by a compactified volume $V_{d-4}$ of extra dimensions 
\begin{align}
   M^2_{Pl} &= V_{d-4} (M_d)^{d-2}\quad.
   \label{eq:int:relation}
\end{align}

The first relation between asymptotic safety to theories including
large extra dimensions was the 
 observation that the NGFP \eqref{NGFP}, providing
the key element of the asymptotic safety program, also exists 
for $d \ge 4$ \cite{Reuter:2001ag,Fischer:2006fz}. Thus asymptotic
safety also constitutes a viable mechanism for theories including extra dimensions.
Subsequently, it was argued that this mechanism could also affect the black hole production process and
its related observables \cite{Koch:2007yt,Hewett:2007st,Litim:2007iu,Burschil:2009va,Contreras:2013hua,Falls:2010he}. 
Further, due to the possibility of the formation of a black hole remnant within asymptotic safety (see figure \ref{figTvonMSS}),
this approach gave a well funded framework for studying the formation of black hole remnants 
at the large hadron collider\ \cite{Koch:2005ks,Bellagamba:2012wz,Alberghi:2013hca}.
This
subsection summarizes the central results of this analysis, mainly following the
notation of Ref.\ \cite{Falls:2010he,Burschil:2009va}.

The starting point is the $d$-dimensional Schwarzschild
solution
\be
ds^2=-f(r)dt^2+ f(r)^{-1} \, dr^2 + r^2 \, d\Omega_{d-2}^2\quad,
\ee
with the radial function
\be
f(r)=1-\frac{2 \, G_N \, M}{r^{d-3}}\, .
\ee
Here $G_N$ denotes the classical, $d$-dimensional Newton's constant, 
the reduced black hole mass $M$ is related its physical mass
 via
\be
M=\frac{4 \Gamma((d-1)/2)}{(d-2)\pi^{(d-3)/2}} \, M_{phys} \, ,
\ee
and the Schwarzschild radius $r_{cl}$ of the solution is given by
\be
r_{cl}^{d-3}=2 \, G_N \, M \, .
\ee

We now adapt the RG-improvement procedure of section \ref{subsecSSS} 
to the generalized Schwarzschild black hole.
After replacing the scale dependent coupling $G_N \rightarrow G_k$ one
has to impose the cutoff identification relating $k$ to the physical scale. The relation (\ref{dproperr})
can be generalized straightforwardly to the $d$-dimensional case, yielding
\be\label{drXD}
d_{r}(r)= \int_0^r dr' \left| 1- \left(\frac{r_{cl}}{r'}\right)^{d-3}\right|^{-1/2} \, .
\ee
For large radii $r\gg r_{cl}$, $d_r(r)$ increases linearly with $r$
\be
\lim_{r \rightarrow \infty} d_{r}(r)=r \, , 
\ee
while for small radii $r \ll r_{cl}$ the asymptotics is given by
\be
\lim_{r \rightarrow 0} d_{r}(r)=\frac{2 r^{(d-1)/2}}{(d-1) r_{cl}^{(d-3)/2}} \, .
\ee
The complete $d_r(r)$ is obtained by numerically integrating (\ref{drXD})
and shows two asymptotic regions, which are connected
by a short transition regime situated at $r \simeq r_{cl}$.
In the spirit of (\ref{interpol}) one can define an interpolating function
that connects these asymptotic scaling regimes
\be\label{interpolXD}
d_{int}(r)=\frac{2 r^{(d-1)/2}}{(d-1) (r_{cl}+\epsilon r)^{(d-3)/2}} \, ,
\ee
Setting $\epsilon= (d/2-1/2)^{-2/(d-3)}$ reproduces the
asymptotic behavior for small and large values $r$ for all $d \ge 4$,
albeit leading to a somewhat more pronounced transition regime.

Combining the interpolating function \eqref{interpolXD} with the
running of $G_k$ obtained in the one-loop approximation \eqref{Gbulkk}
the $r$-dependent Newton's constant becomes
\be
G(r)=\frac{G_0 r^\alpha}{r^\alpha + \tilde \omega G_0 (r_{cl}+ \epsilon r)^{\alpha+2-d}}\quad,
\ee
with
\be
\alpha  =  \frac{1}{2} (d-1)(d-2) \, , \qquad 
\tilde \omega = \omega_d \, \xi^{d-2} \, \left(\frac{d}{2}-\frac{1}{2}\right)^{d-2} \, .
\ee
Finally the improved radial function for the scale setting (\ref{interpolXD}) reads
\be\label{frImpXD}
f(r)= 1-\frac{G_0 r^\alpha}{r^\alpha + \tilde \omega G_0 (r_{cl}+ \epsilon r)^{\alpha+2-d}} \frac{M}{r^{d-3}}\quad.
\ee

\begin{figure}[t!]
\centerline{\includegraphics[width=10cm]{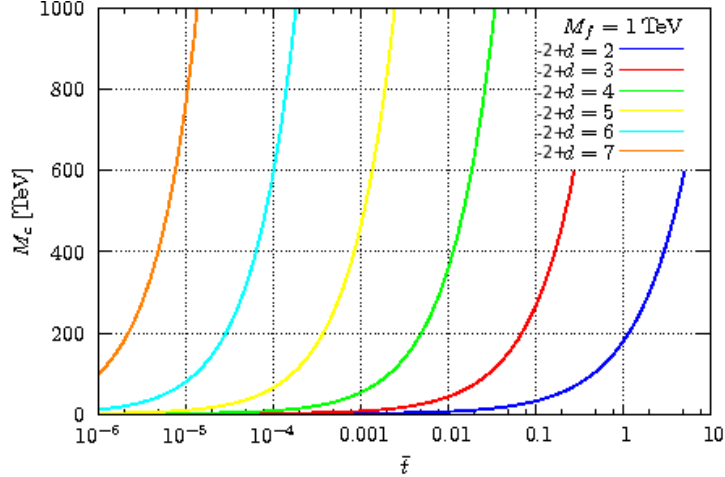}}
\caption{\label{figMcXD} 
Dependence of the critical mass $M_{cr}$ on the parameter $\tilde t$ 
for various numbers of dimensions and a reduced Planck mass of $M_d=1 TeV$. This parameter can be related to the parameter on $\tilde \omega$ by the relation $\tilde t\approx \tilde \omega G_0^2 M_{cr}^2 M_d^{d-2}$\cite{Burschil:2009va}. }
\end{figure}

The asymptotic behavior and horizon structure entailed by the improved radial function
is essentially the same as for the four-dimensional case shown in figure \ref{figfvonrSS}. 
Again there is a critical mass parameter $M_{cr}$. For $M > M_{cr}$ the black hole
possesses an inner and outer horizon while for $M < M_{cr}$ no horizon appears.
For $M = M_{cr}$ the improved radial function has a double zero. From \eqref{HT}
one then concludes that for the critical mass the horizon temperature
vanishes. Thus $M_{cr}$ constitutes the remnant mass, due to the non-radiating property
of the corresponding critical black hole. The dependence of $M_{cr}$
on the parameter $\tilde \omega$ is illustrated in figure \ref{figMcXD}.

The main interest in higher dimensional black holes comes from
the possibility of black hole production due to high energy collision of particles
with invariant collision energy $\sqrt{s}$.
In a semiclassical approximation the cross section for such a black hole production was estimated to
\be\label{crosssecalt}
\sigma(\sqrt{s})= \pi \, r_{cl}^2 \, \theta(\sqrt{s}-M_d)\quad,
\ee
where usually, the minimal mass for this process was assumed to be the reduced 
the Planck mass $M_d$, a condition which is implemented in (\ref{crosssecalt}) 
by the theta function $\theta(...)$.
This approximation of the cross section turned out to also be valid
in different approaches (for a more detailed discussion, we refer to Refs.\ 
\cite{Voloshin:2001fe,Jevicki:2002fq,Eardley:2002re,Rychkov:2004sf,Kang:2004yk,
Rizzo:2006uz}).

\begin{figure}[t!]
\centerline{\includegraphics[width=10cm]{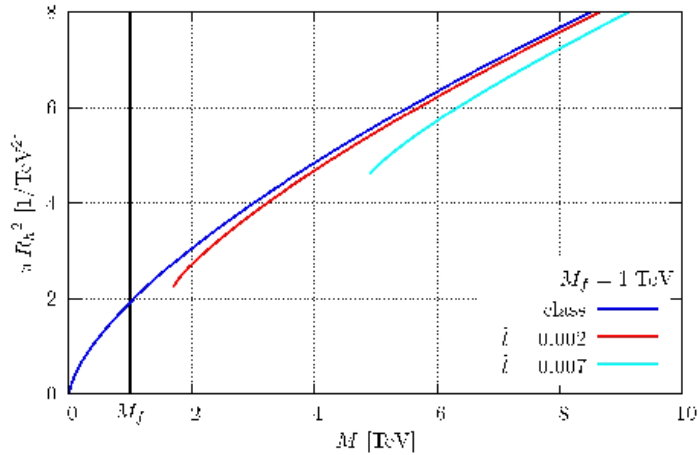}}
\caption{\label{figSigmaXD} 
Black hole production cross section as a function
of $\sqrt{s}$ for $d=6$, $M_d=1$~TeV and for different values of $\tilde t$
\cite{Burschil:2009va}.}
\end{figure}

In the context of asymptotic safety this prediction has to be reexamined.
The first generalization comes with the observation,
that the actual horizon radius of (\ref{frImpXD}) gets shifted towards smaller
values, as it can be seen from figures \ref{figfvonrSS}.
This implies a suppression of the black hole production cross section.
The second generalization comes from the interpretation of the minimal scale
at which this process is expected to occur. Since for masses below $M_{cr}$
there is actually no more horizon one can actually only speak of black hole formation
for $M>M_{cr}$. Thus, the most straight forward black hole production cross section
in the context of asymptotically safe black holes is
\be\label{crosssec}
\tilde \sigma(\sqrt{s})= \pi \, r_+^2 \, \theta(\sqrt{s}-M_{cr})\quad.
\ee
In figure \ref{figSigmaXD} one can see this cross section for various values of $\tilde t\approx \tilde \omega^{-1/d}$.
One observes that for the production of very massive black holes $M\gg M_{cr}$
the improved cross section agrees with the semi-classical estimate.
However, when $M$ approaches $M_{cr}$,
the numerical values start to differ significantly.
The most drastic difference to the semi-classical estimate appears when $M=M_{cr}$,
since this defines the new threshold for black hole production.
One sees that this threshold $M_{cr}$ can be significantly higher than
the ad hoc threshold $M_d$.
Since due to the form of the parton distribution function, the actual production rate for those black holes is strongly peaked
at the threshold this implies that
the predicted cross section (\ref{crosssecalt}) would actually largely
overestimate the semi classical black hole production rate with respect to the
rate calculated from (\ref{crosssec}) \cite{Burschil:2009va}.

\section{Black holes including the cosmological constant}
\label{sect.4}
An important observation based on figure \ref{EHflow}
is that the gravitational RG flow dynamically generates a
cosmological constant even if it is zero at one particular scale.
The scaling dynamics associated with the NGFP \eqref{UVG} shows that
the dimensionful Newton's constant actually flows to zero while the value
of the cosmological constant diverges as $k \rightarrow \infty$. On this basis 
it is natural to investigate the role of the cosmological constant
in the RG-improvement process. The discussion of section \ref{sect.4a} and \ref{sect.4b}
mainly follows Ref.\ \cite{Koch:2013owa} while section \ref{sect.5} contains new material.
%
\subsection{RG improved (A)dS black holes}
\label{sect.4a}
The line element of a static, spherical symmetric black hole
in the presence of a cosmological constant is again of the form \eqref{ssmetric}.
The radial function $f(r)$ now includes an additional
term containing $\Lambda_0$
\be\label{fads}
f(r) = 1 - \frac{2 \, G_0 \, M}{r} - \frac{1}{3} \, \Lambda_0 \, r^2 \, . 
\ee
Similarly to the classical solution this geometry possesses a curvature singularity
at the origin
\be\label{singClass}
R_{\m\n\rho\sigma} R^{\m\n\rho\sigma} = \frac{48 G^2 M^2}{r^6} + \frac{8 \Lambda^2}{3} \, .
\ee
This singularity is shielded by a horizon structure where \eqref{fads} vanishes. This structure
depends on the sign of $\Lambda_0$ and $M$. For $\Lambda_0 \le 0$ there is a single horizon
which agrees with the one of the Schwarzschild black hole $r_{\rm SS} = 2 G_0 M$ for $\Lambda_0 =0$.
For $\Lambda_0 > 0$ and $M < (3 G_0 \sqrt{\Lambda_0})^{-1}$ the geometry possesses two horizons, an inner black hole horizon and an outer cosmological horizon.
For the critical mass $M_{\rm crit} = (3 G_0 \sqrt{\Lambda_0})^{-1}$ the two horizons coincide and the line element
describes the Nariai black hole as the maximal black hole in de Sitter (dS) space. For $M > M_{\rm crit}$ there is no horizon
and the geometry possesses a naked singularity.

\begin{figure}[t]
  \centering
\includegraphics[width=0.48\textwidth]{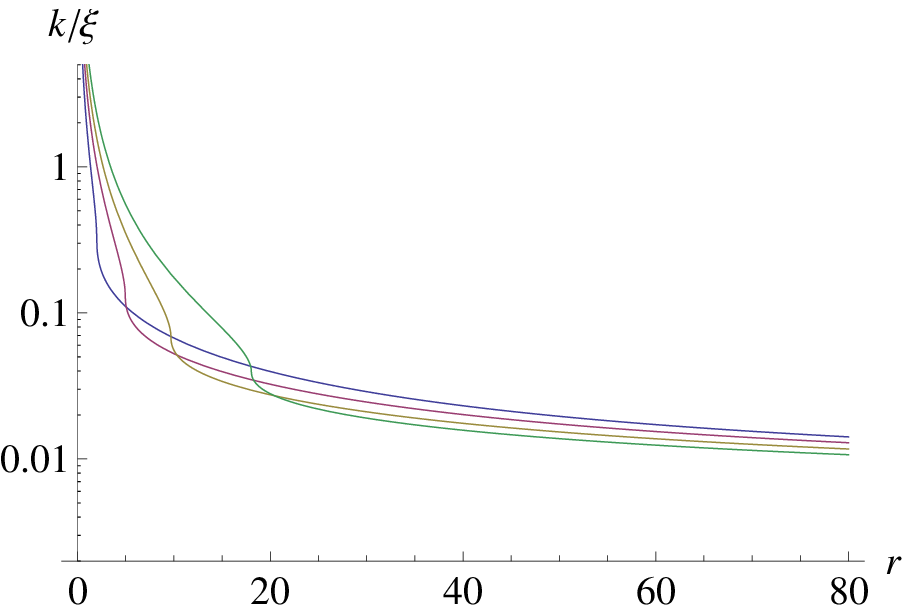}
\hspace{0.1cm}
\includegraphics[width=0.48\textwidth]{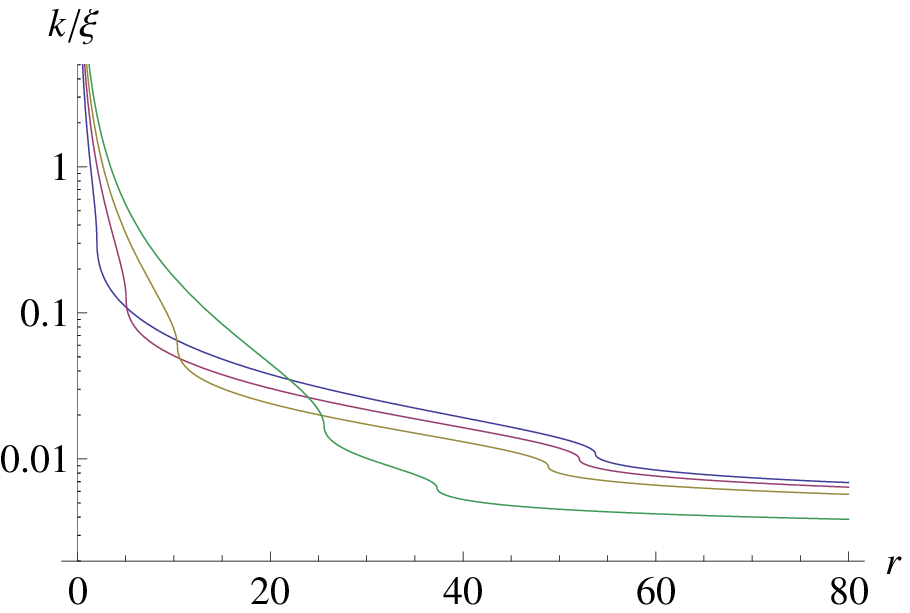}
\caption{Cutoff-identification $k(r)$ resulting from identifying $d(r)$ with the radial geodesic distance \eqref{dproperr}. The left and right diagram
illustrate the function $k(r)/\xi$ for $G_0 = 1$ and $M=\{1, \,2.5, \,5, \, 10 \}$ (top to bottom curve) for the case of a Schwarzschild-AdS black holes with $\Lambda_0=-0.001$
and a Schwarzschild-dS black holes with $\Lambda_0=0.001$, respectively. \label{gl2}}
\end{figure}

The construction of the RG improved (A)dS black hole directly follows
the Schwarzschild case. The first step promotes the coupling constants
appearing in the classical expression \eqref{fads} to scale-dependent
quantities
\be\label{fadsk}
f_k(r) = 1 - \frac{2 \, G_k \, M}{r} - \frac{1}{3} \, \Lambda_k \, r^2 \, . 
\ee
The cutoff identification relating $k$ to the radial distance $r$, $d(r)$ can be constructed numerically. For selected sample black holes
the corresponding identification arising from \eqref{dproperr} is shown in figure \ref{gl2}. The  proper time improvement \eqref{dpropert} gives rise to a quite similar identification, which has the benefit that is is smooth at the classical horizons.
The short distance behavior of $d(r)$ is again governed by the $G_0$-term in \eqref{fads} so that the expansion
\eqref{d1inin} carries over to the Schwarzschild-(A)dS geometry.

We now use the RG improvement procedure to ``zoom into'' the classical black hole singularity. Being interested in
the short distance or high energy limit, we start from the classical radial function \eqref{fads}, promote the coupling constants
to scale-dependent quantities and subsequently describe the running of $\Lambda_k, G_k$ by the scaling laws entailed by
the NGFP. The RG improved radial line element $f_*(r)$ obtained in this way reads
\be\label{fvonrimp}
f_*(r) = 1-\frac{2\, M \, g_*}{k^2 \, r} - \frac{1}{3} \left( \lambda_* \, k^2 \right) r^2 \, . 
\ee 
Replacing $k(r)$ by the leading term of the asymptotics \eqref{krasym} yields the RG-improved line element
valid at the NGFP
\be\label{fself}
f_{*}(r) = 1-  \frac{2\, G_0 \, M}{r} \left( \frac{3}{4} \, \lambda_* \, \xi^2 \right) - \frac{1}{3} \, \left( \frac{4 \, g_*}{3 \, G_0 \, \xi^2} \right) \, r^2 \, . 
\ee   
Most remarkably, the RG-improved function $f_{*}(r)$ is \emph{self-similar} to the classical solution \eqref{fads}. Promoting $k$ to a function
of $r$ interchanges the terms containing Newton's constant and the cosmological constant so that the actual $r$-dependence remains the same.
This entails in particular that the RG-improved line element also gives rise to a curvature singularity with precisely the same structure
as in the classical case. In this sense the inclusion of the cosmological constant reverted the resolution of the classical singularity
observed in the Schwarzschild case.

Based on \eqref{fself}, one can also identify a special value for the a priori undetermined constant $\xi$. Adopting
\be\label{xisc}
\xi^2_{\rm sc} = \frac{4}{3 \lambda_*} \, .
\ee
the RG-improvement scheme becomes self-consistent in the sense that both the classical and the RG-improved line element give
rise to the same cutoff identification at short distance. For this value the radial function
describes a Schwarzschild-dS black hole with effective cosmological constant
\be
\Lambda_{\rm eff} = \frac{g_* \, \lambda_*}{G_0} \, . 
\ee
Notably, the dimensionality of $\Lambda_{\rm eff}$ (valid in the UV) is set by the square of the Planck mass $M_{\rm Pl}^2 \equiv G_0^{-1}$, while its magnitude (in Planck units) is governed by the universal dimensionless product $g_* \, \lambda_*$. This product cannot be chosen by hand, but constitutes a prediction from Asymptotic Safety. Its magnitude has been computed in a number of works 
$g_* \, \lambda_* \approx 0.1$ \cite{Reuter:2012id,Reuter:2012xf}.

\subsection{Hawking radiation and black hole evaporation}
\label{sect.4b}
\begin{figure}[t]
  \centering
\includegraphics[width=0.48\textwidth]{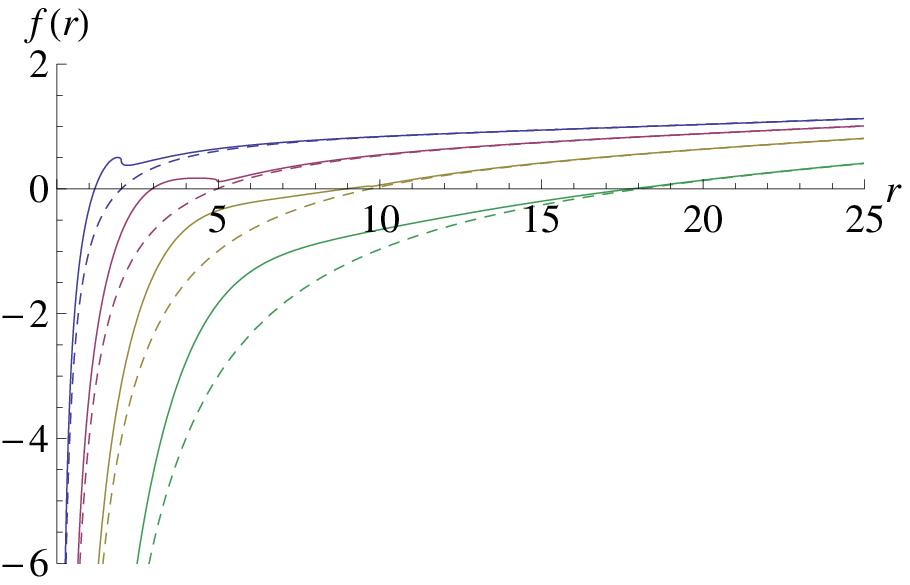}
\hspace{0.1cm}
\includegraphics[width=0.48\textwidth]{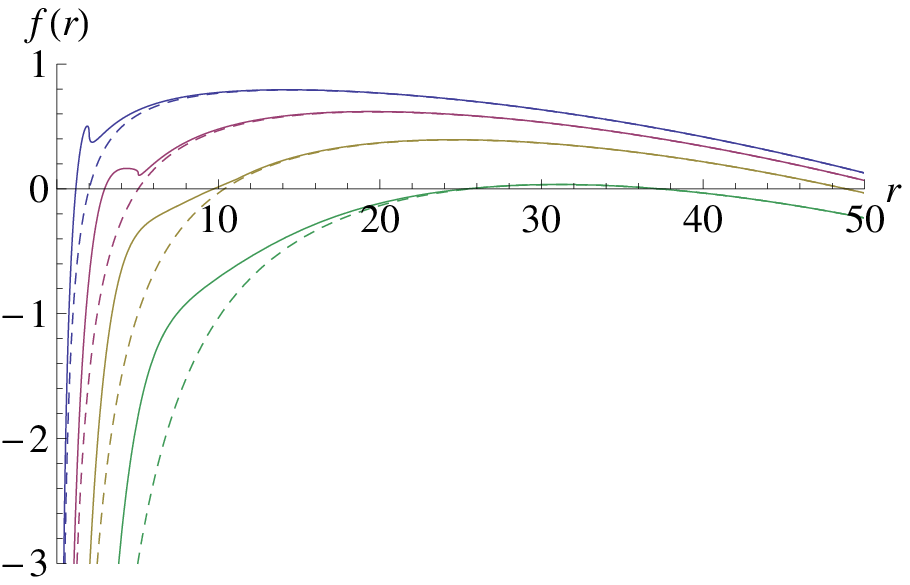}
\caption{Radial dependence of the RG-improved metric function $f(r)$ for $\xi = \xi_{\rm sc}$.   The dashed lines correspond to the classical solutions while the solid lines
 correspond to the improved solution obtained from $G_0=1$, and
 from top to bottom $M=\{1,\, 2.5, \,5, \, 10\}$ and $\Lambda_0=-0.001$ (left panel) and $\Lambda_0=0.001$ (right panel).
\label{gl3}
}
\end{figure}

The complete $r$-dependence of the RG-improved radial function
is found by modeling the $k$-dependence of the coupling constants 
appearing in \eqref{fadsk} by 
the analytic curves \eqref{lvong} and subsequently replacing
$k \mapsto k(r)$ with the functions $k(r)$ shown in figure \ref{gl2}.
The comparison between the RG-improved radial function and its classical counterpart
is shown in figure \ref{gl3}.
Comparing these curves to the ones shown in figure \ref{figfvonrSS} establishes that the inclusion of the cosmological constant changes the short-distance behavior of the improved line element,
so that it is again similar to the classical version.

\begin{figure}[t!]
\centerline{\includegraphics[width=8cm]{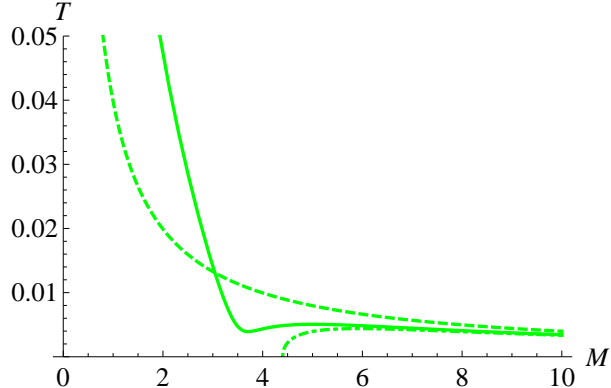}}
\caption{Comparison between the temperature of the (RG-improved) Schwarzschild
black hole with $\Lambda_0 = 0$. The dash-dotted (lower) curve was obtained in
\cite{Bonanno:2000ep} by setting $\Lambda_k = 0$ for all values $k$ while the solid curve
includes the RG-running of $\Lambda_k$. The classical behavior is given by 
the dashed upper curve for comparison. From Ref.\ \cite{Koch:2013owa}. \label{fig.1}}
\end{figure}

The Hawking temperature associated with the inner black hole horizon is found by evaluating
\eqref{HT} for the RG-improved $f(r)$ as a function of the mass parameter $M$. 
For the asymptotically Schwarzschild black hole with $\Lambda_0 = 0$ the result
is shown as the solid curve of figure \ref{fig.1}. For comparison the corresponding
temperature for the classical Schwarzschild black hole is shown as the dashed curve
while the dashed dotted curve shows the RG-improved Schwarzschild black hole 
without including the cosmological constant discussed in section \ref{subsecSSS}.
As its most important feature, the inclusion of $\Lambda_k$ prevents that the
temperature of the inner horizon drops to zero at a finite mass $M_{cr}$. Thus it
is expected that the RG-improved black hole solution evaporates completely.
In this analysis there is no formation of Planck-mass black hole remnants within
asymptotic safety.

\subsection{Tracing the origin of the quantum black hole singularity}
\label{sect.5}
A rather surprising feature of the RG-improvement procedure is
the observation that including the cosmological constant in the improvement process reintroduces
the black hole singularity, that was removed in section \ref{sect.3}. Even though the classical
and quantum improved black hole singularity are formally of the same form, their physical origin
is actually quite different. This is demonstrated by applying the RG-improvement process
to empty space.

The starting point is again the line element \eqref{ssmetric}. For empty space, including
a cosmological constant, the classical radial function is given by
\be
f(r) = 1 - \frac{1}{3} \, \Lambda_0 \, r^2 \, .
\ee
For $\Lambda_0 = 0$, $\Lambda_0 > 0$ and $\Lambda_0 < 0$, this line element provides 
a metric on flat Minkowski space, de Sitter space, and Anti-de Sitter space, respectively.
These spaces are maximally symmetric which in particular implies that they are homogeneous, so that 
``all points are the same''. We then select the point $r=0$ and use the RG improvement
to zoom in on this particular point. 

For concreteness, the classical spacetime is chosen to be flat Minkowski space with $\Lambda_0 = 0$.
Adapting the cutoff identification \eqref{kvonP} to this case yields
\be
k(r) = \frac{\xi}{r} \, . 
\ee
This identification attaches a double meaning to $r$
which acts as a coordinate and, at the same time, 
indicates the linear volume size over which the RG improved quantities are averaged.
\begin{figure}[t!]
\centerline{\includegraphics[width=8cm]{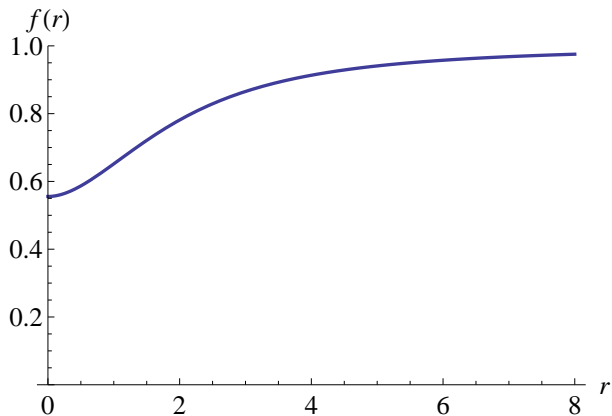}}
\caption{The function $f(r)$ describing the line element for RG-improved flat space, $\Lambda_0 = 0$ with the scale-identification $k = \xi_{\rm sc}/r$. For $r \gg 1$, $f(r)$ approaches its flatspace limit $f(r) = 1$, while for $r \rightarrow 0$, $f(r) \simeq 0.55$ becomes constant. \label{fig.5}}
\end{figure}
Introducing the scale-dependence of the cosmological constant $\Lambda_0 \mapsto \Lambda_k = \lambda_k \, k^2$ with the running of $\lambda_k$ governed by
\eqref{lvong}, the cutoff identification gives rise to the RG improved radial function shown in figure \ref{fig.5}.
The function smoothly interpolates between classical flat space and a improvement scheme dependent positive constant $f(0) < 1$. 
Notably $f(r)$ remains positive throughout so that the RG-improvement process does not lead to the formation of a horizon. 
At this stage, it is illustrative to investigate the square of the Riemann tensor, obtained from the
full RG improved line element. This quantity is shown in figure \ref{fig.6} and interpolates
smoothly between vanishing curvature for $r \gg 1$ and a scalar curvature singularity at $r=0$.
\begin{figure}[t!]
\centerline{\includegraphics[width=8cm]{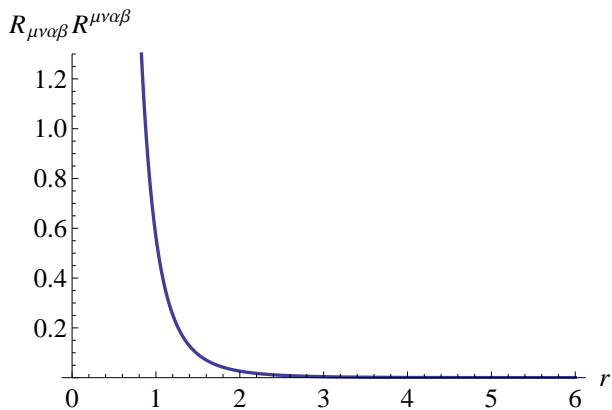}}
\caption{Square of the Riemann curvature tensor as function of the coarse-graining distance $r$. For $r \gg 1$ we obtain flat space. For $r \rightarrow 0$ the squared curvature diverges, indicating the development of a scalar curvature singularity. \label{fig.6}}
\end{figure}
Remarkably, RG improving flat space including a cosmological constant also gives rise to a singularity. Thus
the singularity observed in the last section also appears in the absence of a black hole. Therefore it is important to carefully distinguish between 
the classical black hole singularity and the singularity appearing for the quantum improved black hole, since
these two effects may come from a very different physics origin.

The features of this singularity can be understood from
investigating the $r \rightarrow 0$ (or, equivalently, the $k \rightarrow \infty$) limit. In this limit
the scale dependence of the cosmological constant is controlled by the NGFP, eq.\ \eqref{UVG}.
Substituting this scaling relation the improved line element becomes
\be
ds^2_* = - \alpha \, dt^2 + \alpha^{-1} \, dr^2 + r^2 \, d \Omega_2^2 \, . 
\ee
where
\be
\alpha = 1 - \frac{1}{3} \, \lambda_* \, \xi^2 \, , 
\ee
is positive as long as $\xi^2 < 3/\lambda_*$. This line element actually
describes a \emph{conical singularity} at $r = 0$. Computing the square of 
the Riemann tensor based on the line element, one finds a scalar curvature singularity
at $r = 0$:\footnote{A similar behavior is expected from Regge calculus, where the
discretized curvature scalar also diverges in the continuum limit \cite{}.}
\be\label{singbe}
R_{\m\n\alpha \beta} R^{\m\n\alpha \beta} =  \frac{4 (\alpha -1)^2}{r^4} \, . 
\ee
At the same time all components of $R_{\m\n\alpha \beta}$ and the Ricci tensor
remain finite at $r=0$, so that an observer experiencing this singularity
is not subject to diverging tidal forces. 
Also geodesics do not experience 
singular behavior when passing $r=0$. Thus, the singularity is of
a much milder form that the one encountered in the black hole case.
This is readily understood from the fact that in empty space, $r$ is the only quantity which carries a unit, so
conformal invariance of the fixed point theory fixes \eqref{singbe}.
Based on on this observation we expect that the singular behavior of the RG improved black hole rather reflects a microscopic feature 
of the quantum spacetime than being related to black hole physics. It is tempting to speculate 
that the singular behavior of the asymptotically safe quantum spacetime is actually
related with the feature of dynamical dimensional reduction, Ref.\ \cite{Carlip:2009kf}, along 
the ideas advocated in Refs.\ \cite{Futamase:1984ej,Hu:1986cv,Berkin:1992ub}.

\section{Further remarks}
\label{sect.7}
We conclude our review by briefly commenting on further developments
contributing to the understanding of black holes in asymptotically safe gravity. 
%
\subsubsection*{The effect of higher-derivative terms}
%
So far, the discussion of the quantum structure of black holes emerging within asymptotic safety approximated the effective average action $\Gamma_k$ by the Einstein-Hilbert truncation \eqref{action} including one ($G_k$) or two ($G_k$ and $\Lambda_k$) running couplings. Besides these terms, $\Gamma_k$ will contain higher-derivative terms (see Refs.\ \cite{mr,Lauscher:2001rz,Codello:2007bd,Machado:2007ea,Benedetti:2009rx,Benedetti:2010nr}) 
or terms of a ``bi-metric'' nature \cite{Manrique:2010mq,Manrique:2010am}. The effect of higher-derivative terms on the RG improvement of the classical black hole solution has been investigated in Ref.\ \cite{Cai:2010zh}, which considered the analogue of the Schwarzschild-(A)dS black hole solutions in higher-derivative gravity. While the incorporation of higher-derivative terms gives rise to a richer class of solutions, one can choose boundary conditions that the line element is of the form \eqref{fads} where the cosmological constant becomes an effective cosmological constant receiving contributions from the higher-derivative couplings.

A conceptual difference of this work to asymptotic safety is that that the RG flow capturing the scale dependence of the higher-derivative couplings is based on the perturbative beta-functions of higher-derivative gravity \cite{Julve:1978xn,Salam:1978fd} incorporating the non-Gaussian fixed point values for Newton's constant and the cosmological constant \cite{Codello:2006in,Groh:2011vn}. In this case the higher-derivative couplings logarithmically flow to zero at high energies, which does not reflect the flow to a non-Gaussian fixed point as expected from non-perturbative calculations \cite{Benedetti:2009rx}.

The cutoff-identification implemented in Ref.\ \cite{Cai:2010zh} is based on solving RG-improved equation of motion. Imposing suitable boundary conditions, the resulting cutoff identification for large values $k$ has the asymptotic expansion
\be\label{Caiimp}
k(r) \propto r^{-3/4} \, . 
\ee
Substituting the result in the scale-dependent line element, the radial function at short distances becomes
\be
f_*(r) = 1 - c \sqrt{r} \, , 
\ee
where the constant $c$ depends on the fixed point values of the coupling constants. The RG-improved black hole solutions found in this way support a picture very similar to the one encountered in asymptotic safety:
\begin{itemize}
\item For large values $r \gg l_{\rm Pl}$ solutions agree with the classical Schwarzschild-(A)dS solution.
\item The curvature singularity at the center of the black hole is weaker but contrary to the Schwarzschild case not resolved completely.\footnote{In Ref.\ \cite{Cai:2010zh} this singular behavior was attributed to the inclusion of the higher-derivative terms. Following the computation in the light of the findings presented in the last section, it is found that the effect also appears in the absence of the higher-derivative terms and can be tracked  to the inclusion of the cosmological constant.}
\item The RG improvement leads to the appearance of an event horizon and a Cauchy horizon, changing the asymptotics of $f(r)$ at short distances. As the mass of the black hole is lowered, the two horizons approach each other and the coincide for a critical mass $M_{cr} > 0$. 
\item 
Once the black hole attains critical mass its horizon temperature becomes zero. Thus the RG-improvement scheme based on \eqref{Caiimp} supports the formation of cold Planck mass black hole remnants.
\end{itemize}

\subsubsection*{State counting based on $\Gamma_k$}
A central goal any quantum theory of gravity should strive for is a 
microscopic explanation of the Bekenstein-Hawking entropy \eqref{BHent}
in terms of fundamental degrees of freedom. The natural interpretation
of this entropy within the context of the effective average action is clear:
the entropy arises from the statistical mechanics of the {\it geometrical fluctuations} about their respective 
background. 
In Ref.\ \cite{Becker:2012js} this proposal was made precise
by proposing a state counting formula based on $\Gamma_k$.

This procedure is based on self-consistent backgrounds which arise as solutions of the field equations
obtained when varying $\Gamma_k$ with respect to the fluctuation fields\footnote{In the context of black holes and the AdS-CFT correspondence similar variational
methods were used to properly regularize and renormalize black hole solutions \cite{Andrade:2006pg}.}
\be\label{defselfcon}
\left. \frac{\delta}{\delta h_{\m\n}} \, \Gamma_k\left[h; \bar{g}_k^{\rm selfcon} \right] \right|_{h = 0} = 0 \, . 
\ee 
The relation (\ref{defselfcon}) implies the expectation value of the fluctuation field, $h_{\m\n}$ vanishes for the chosen background.
For the ``single-metric'' ans\"atze discussed in section \ref{sect.2} this boils down to the background satisfying the
scale-dependent equations of motion obtained from $\Gamma_k$.\footnote{In the context of ``bi-metric'' truncations, \eqref{defselfcon} implies that $\bar{g}_k^{\rm selfcon}$ solves the tadpole equation. We refer to Ref.\ \cite{Becker:2012js} for a detailed discussion of this point.}

Starting from the exact integro-differential equation satisfied by $\Gamma_k$, Ref.\ \cite{mr}, one can obtain
the following path integral representation of $\Gamma_k[0, \bar{g}_k^{\rm selfcon}]$
\be
\mathbb{Z}_k \equiv e^{-\Gamma_k[0, \bar{g}_k^{\rm selfcon}]} = \int \cD \hat{\Phi} \, e^{-\tilde{S}[\hat{\Phi}, \bar{g}_k^{\rm selfcon}]} \, e^{-\Delta_kS[\hat{\Phi}]} \, . 
\ee
Here $\hat{\Phi}$ denotes the fluctuation fields with vanishing fluctuation averages $\langle \hat{\phi} \rangle = 0$ around the
background $\bar{g}_k^{\rm selfcon}$, $\tilde{S}$ is the gravitational action supplemented by suitable gauge-fixing and ghost 
terms, and $\Delta_kS[\hat{\Phi}] \sim \int d^dx \sqrt{\gb} \, \hat \phi \,  \cR_k \, \hat \phi$ is an IR regulator providing a mass-type cutoff for fluctuations with momentum $p^2 < k^2$.
The r.h.s.\ thus resembles a partition function of a statistical system with action $\tilde{S}$ cut off at the IR scale $k$. 
In this sense $\mathbb{Z}_k$ provides a tool for `counting' the states (field modes) integrated out between infinity and the IR scale $k$.

It is illustrative to apply this state-counting formula to the Euclidean Schwarzschild solution
\be\label{EuclSS}
ds^2 = f(r) dt^2 + f(r)^{-1} dr^2 + r^2 d \Omega^2_2
\ee
with 
\be\label{BRline}
f(r) = 1- \frac{r_s}{r}
\ee
and $r \in [r_s, \infty]$ and the time coordinate periodic with $t \in [0, \beta]$ where $\beta = 4 \pi r_s$ is the
inverse temperature of the black hole. This spacetime manifold $\cM$ has two boundaries, the horizon at the Schwarzschild radius $r_s$
and at infinity. In contrast to the ``improved solution'' techniques, where $r_s = 2 M G_k$ is a scale-dependent 
quantity, the $r_s$ appearing in \eqref{BRline} is treated as a \emph{scale-independent} integration constant
labeling the static and spherical symmetric solutions of \eqref{defselfcon}. 

The $\mathbb{Z}_k$ for the Schwarzschild solution has been obtained in Ref.\ \cite{Becker:2012js} and 
the extension to the Nariai black hole has been carried out in Ref.\ \cite{Koch:2013owa}. In the Schwarzschild case,
 the solution \eqref{EuclSS} is Ricci-flat, $R_{\m\n} = 0$, the only contribution
to $\mathbb{Z}_k$ comes from the boundary terms
\be
\begin{split}
- \ln \mathbb{Z}_k = & \, \Gamma_k[\bar{h} = 0, \bar{g}_{\rm SS}] \, \\
= & - \frac{2}{16 \pi G_k^\p} \, \int_{\p \cM} d^3x \sqrt{\bar{H}}  \left( \bar{K} - \bar{K}_0 \right) -   \frac{2}{16 \pi G_k^\p} \, \int_{\p \cM} d^3x \sqrt{\bar{H}}  \bar{K}_0 \, . 
\end{split}
\ee
Here the first term is the usual Gibbons-Hawking term, while the second term vanishes to leading order in $r_s/r$. Since
the second term encodes the surface at infinity, is independent of the spacetime curvature caused by the black hole and just encodes
properties of flat space. Thus it will be neglected. The evaluation of the first integral then yields
\be
- \ln \mathbb{Z}_k = \frac{\beta r_s}{4 G^\p_k} + \ldots = \frac{A}{4 G^\p_k} + \ldots \,  .
\ee
where $A \equiv 4 \pi r_s^2$ is the area of the (Euclidean) event horizon. Thus for an RG trajectory where $\lim_{k \rightarrow 0} G^\p_k = G_0$ in the IR
the formalism recovers the Bekenstein-Hawking area law as the leading term. Moreover,
the formalism can be extended to compute the corrections to the area law in a systematic way.

\subsubsection*{The scale-dependent ADM mass}
Using the Schwarzschild radius $r_s$ as the quantity that sets 
the typical length scale of the (Euclidean) Schwarzschild geometry,
one can introduce the scale-dependent ADM mass
\be\label{ADMmass}
M_k \equiv \frac{r_s}{2 G_k^\p} \, . 
\ee
Note that the effective average action formalism actually
associates the {\it boundary Newton's constant} with the mass
of the geometry. Using the flow of $G_k^\p$ from \eqref{Gboundk}
and expanding the result for small values $k$ one finds
\be
M_k = \left[ 1 - |\omega_4^\p| \tfrac{k^2}{G_0^\p} \right] \, M_0 \, .
\ee
Thus $M_k$ actually decreases for increasing values $k$ and reaches zero
near the Planck scale, set by the IR-value of $G_0^\p$. In Ref.\ \cite{Becker:2012js}
it was argued that this behavior agrees with the expectation of gravitational anti-screening: 
as a consequence of the definition \eqref{ADMmass}
the running mass of any material body decreases with increasing $k$, or decreasing distance,
in agreement with the picture of a cloud of virtual excitations surrounding the massive body
put forward in section \ref{sect.2}.

\section{Conclusions and outlook}
\label{sect.6}
A crucial benchmark test for any quantum theory of gravity
is the development of a consistent quantum picture of black holes.
This review summarizes the understanding of black hole physics within
the gravitational asymptotic safety program (QEG, see\ \cite{Reuter:2012id}).
While the present picture is far from being complete, renormalization group improvement methods have already provided
important insights on the leading quantum corrections to classical black hole physics expected from asymptotic safety.
In particular, there is the need of carefully distinguishing between the cases where the improvement process includes the effect of a running cosmological constant or not,
since the two setups shine spotlights on very different physical phenomena
and drastically change the structure of black holes at subplanckian scales. 

The picture of the renormalization group improved Schwarzschild black hole solution
has essentially been developed in Refs.\ \cite{Bonanno:2000ep,Bonanno:2006eu}. For distances
$r \gg 2 G_0 M$ the quantum-improved solutions  coincide with their classical counterparts.
For $r \ll 2 G_0 M$ the improvement procedure drastically alters the structure of the classical
black hole, thereby confirming the common expectation that the RG-improvement process resolves the black hole singularity, 
replacing the black hole interior by a smooth de Sitter like spacetime patch.
Moreover, the RG improvement alters the horizon structure of the improved black hole solution
such that the semi-classical black hole evaporation process leads to 
the formation of a Planck mass cold remnant. Similar pictures
supporting the formation of a black hole remnant and the softening
of the black hole singularity have been derived from
an RG improvement (based on one-loop beta functions) including higher-derivative terms in Ref.\ \cite{Cai:2010zh}
and the analysis of the RG improved from a thermodynamical viewpoint \cite{Falls:2012nd}.

This picture changes drastically by extending the analysis to the class of asymptotically (A)dS black holes \cite{Koch:2013owa,Koch:2013rwa}
 taking the scale-dependent cosmological constant into account. 
 The flow diagram shown in figure \ref{EHflow} thereby illustrates that the 
 RG flow generates a cosmological constant, even if it was not present in the classical low energy action.
  Remarkably, it is the running cosmological constant that dominates the short distance behavior of the RG-improved black hole.
 Including $\Lambda$ reintroduces the classical black hole singularity that was resolved in the case of the RG-improved Schwarzschild solution.
 The improved black hole solution obtained at the NGFP has precisely the same form as its classical counterpart albeight with
 the role of the terms including Newton's constant and the cosmological constant interchanged. This picture
 calls into question the previous conclusion that black hole evaporation in asymptotic safety produces black hole remnants,
 as the RG-improved light black hole would evaporate in a same way as its classical counterpart.

The recurrence of a singularity in the RG-improved black hole  is rather counterintuitive from a quantum gravity point of view: 
after all one motivation for developing 
a theory of quantum gravity stems from the desire to find a resolution of the singularities appearing within classical general relativity.
Notably, the application of the RG improvement methods to flat space, for the first time carried out in section \ref{sect.5},
introduces a similar singularity through the improvement process. This makes it
conceivable, that the ``quantum singularity'' has no direct connection to black hole physics
but reflects a more general phenomenon. Given the rather mild singularity appearing
in the RG improvement process makes it tempting to speculate that this feature realizes
the dynamical dimensional reduction of asymptotically safe quantum spacetime, which has been previously observed 
based on the graviton propagator \cite{Lauscher:2001ya} and later on refined using the spectral dimension 
\cite{Lauscher:2005qz,Reuter:2011ah,Rechenberger:2012pm,Calcagni:2013vsa} at the level of the quantum improved geometry. 

At this stage it is clear that the investigation of black holes within the gravitational
asymptotic safety program is still at its beginning. 
The results obtained so far rely heavily on RG improvement techniques applied to classical solutions 
and on Hawking's semi-classical analysis of the thermodynamical properties of black holes in curved spacetime.
Even at this level it is already clear that the physics of black holes within asymptotic safety is 
much richer than initially thought and probably closely intertwined with
the microscopic structure of spacetime itself. 
Unraveling dynamical questions concerning the
dynamical evaporation of Planck size black holes or shedding light on the microscopic origin of the black hole
entropy will certainly require new techniques that go beyond RG improvements. 

After all, from the experience with the Uehling potential, Ref.\ \cite{Dittrich}, in the context of QED,
 one expects the improvement techniques to work best close to the classical regime.
The transition to the classical regime is in the improvement techniques reflected by
the fact that for large radius $r$ the improved black holes become indistinguishable from their
classical counterparts. 
The improvement techniques are, however,
expected to be less reliable in the 
deep quantum regime (small $r$).
Moreover, since black holes involve more than one typical dimensionful scale
(e.g., the mass of the black hole and the distance from its center) leads to
a certain freedom in identifying the RG scale with a physical scale. 
The uncertainty introduced by the ambiguity of the scale setting procedure
may lead to different predictions 
for small values of $r$. 
A natural next step towards refining the present picture will have to include the interaction between gravitational and matter sector 
in order to verify to what extend the semi-classical analysis based on quantum field theory in a curved spacetime carries over
to the asymptotically safe quantum black holes. We expect that
the gravitational effective average action $\Gamma_k$ together with its functional renormalization
group equation constructed in Ref.\ \cite{mr} will continue to provide guidance for 
addressing these exciting questions in the future.

\section*{Acknowledgments}
We thank H.\ Gies, A.\ Bonanno, M.\ Reuter for helpful discussions. Furthermore, we thank M.\ Reuter for the kind 
permission of reproducing figure \ref{figPenroseSS} from the original Ref.\ \cite{Bonanno:2000ep}
 and figures \ref{figdrKerr} and \ref{figHorizonsKerr} from Ref.\
\cite{Reuter:2010xb}. The work of B.K.\ was supported proj.\ Fondecyt 1120360 
and anillo Atlas Andino 10201 while the research of F.S.\ is
supported by the Deutsche Forschungsgemeinschaft (DFG)
within the Emmy-Noether program (Grant SA/1975 1-1).



\begin{thebibliography}{00}    

\bibitem{Carroll:2004st}
  S.~M.~Carroll,
  {\it Spacetime and geometry: An introduction to general relativity},
  (Addison-Wesley, San Francisco, 2004).
  
\bibitem{Wald:1995yp} 
  R.~M.~Wald,
  {\it Quantum field theory in curved space-time and black hole thermodynamics},
  (Chicago Univ.\ Press, Chicago, 1994).

\bibitem{Taylor} E.\ F.\ Taylor, J.\ A.\ Wheeler,
\textit{Exploring Black Holes, Introduction to General Relativity}, (Addison-Wesley, San Francisco, 2000).

\bibitem{Poisson} E.\ Poisson, \textit{A Relativist's Toolkit}, (Cambridge Univ.\ Press, Cambridge, 2004).


\bibitem{schroeder:nature}
R.\ Sch\"odel, et al.,
Nature {\bf 419.6908}, (2002) 694.

\bibitem{Carr:2009jm} 
  B.~J.~Carr, K.~Kohri, Y.~Sendouda and J.~'i.~Yokoyama,
  Phys.\ Rev.\ D {\bf 81}, 104019 (2010),
  arXiv:0912.5297.



\bibitem{Niedermaier:2006wt} 
  M.~Niedermaier and M.~Reuter,
  {\it Living Rev.\ Rel.\ } {\bf 9}, 5 (2006).

\bibitem{Reuter:2007rv} 
M.~Reuter and F.~Saueressig, in {\it Geometric and Topological Methods for Quantum Field Theory}, H.~Ocampo, S.~Paycha and
 A.~Vargas (Eds.), Cambridge Univ.\ Press, Cambridge (2010), arXiv:0708.1317.

\bibitem{Percacci:2007sz}
  R.~Percacci, in {\it Approaches to Quantum Gravity: Towards a New Understanding of Space, Time and Matter}, D. Oriti (Ed.), Cambridge Univ.\ Press, Cambridge (2009),
  arXiv:0709.3851.

\bibitem{Litim:2008tt}
  D.~F.~Litim, 
  {\it PoS(QG-Ph)} {\bf 024} (2008),
  arXiv:0810.3675.

\bibitem{Reuter:2012id} 
  M.~Reuter and F.~Saueressig,
  {\it New J.\ Phys.\ } {\bf 14}, 055022 (2012),
  arXiv:1202.2274.


\bibitem{Bena:2013dka} 
  I.~Bena and N.~P.~Warner,
  arXiv:1311.4538.


\bibitem{FrancescaBook}
  C.~Rovelli and F.~Vidotto,
  {\it Introduction to Covariant Loop Quantum Gravity},
  (Cambridge Univ.\ Press, Cambridge, 2014).



\bibitem{Banados:1992wn} 
  M.~Banados, C.~Teitelboim and J.~Zanelli,
  {\it Phys.\ Rev.\ Lett.\ } {\bf 69}, 1849 (1992),
  hep-th/9204099.

\bibitem{Emparan:2001wn} 
  R.~Emparan and H.~S.~Reall,
  {\it Phys.\ Rev.\ Lett.\ } {\bf 88}, 101101 (2002),
  hep-th/0110260.

\bibitem{Hawking:1974sw} 
  S.~W.~Hawking,
  Commun.\ Math.\ Phys.\  {\bf 43}, 199 (1975)
  [Erratum-ibid.\  {\bf 46}, 206 (1976)].

\bibitem{Kiefer:2012boa} 
  C.~Kiefer,
  {\it Quantum gravity}, 3rd edition,
  (Oxford Science Publications, Oxford, 2012).



\bibitem{Weinberg:1979}
S.~Weinberg
in \textit{General Relativity, an Einstein Centenary Survey},
S.W.~Hawking and W.~Israel (Eds.),
(Cambridge Univ.\ Press, 1979); 
S.~Weinberg, hep-th/9702027.



\bibitem{mr}
M.~Reuter,
{\it Phys.\ Rev.\ D} {\bf 57}, 971 (1998), hep-th/9605030.



\bibitem{Bonanno:1998ye}
  A.~Bonanno and M.~Reuter,
  {\it Phys.\ Rev.\  D} {\bf 60}, 084011 (1999),
  gr-qc/9811026.

\bibitem{Bonanno:2000ep}
  A.~Bonanno and M.~Reuter,
  {\it Phys.\ Rev.\  D} {\bf 62}, 043008 (2000),
  hep-th/0002196.
  
\bibitem{Bonanno:2006eu} 
  A.~Bonanno and M.~Reuter,
  {\it Phys.\ Rev.\ D} {\bf 73}, 083005 (2006),
  hep-th/0602159.


\bibitem{Reuter:2006rg} 
  M.~Reuter and E.~Tuiran,
  hep-th/0612037.

\bibitem{Reuter:2010xb}
  M.~Reuter and E.~Tuiran,
  {\it Phys.\ Rev.\ D} {\bf 83}, 044041 (2011),
  arXiv:1009.3528.
  

\bibitem{Falls:2012nd} 
  K.~Falls and D.~F.~Litim,
  arXiv:1212.1821.
  

\bibitem{Cai:2010zh} 
  Y.-F.~Cai and D.~A.~Easson,
  {\it JCAP} {\bf 1009}, 002 (2010),
  arXiv:1007.1317.

\bibitem{Becker:2012js} 
  D.~Becker and M.~Reuter,
  {\it JHEP} {\bf 1207}, 172 (2012), arXiv:1205.3583.

\bibitem{Becker:2012jx} 
  D.~Becker and M.~Reuter,
  arXiv:1212.4274.

\bibitem{Koch:2013owa}
  B.~Koch and F.~Saueressig,
  {\it Class.\ Quant.\ Grav.\ }  {\bf 31}, 015006 (2013),
  arXiv:1306.1546.

\bibitem{Koch:2013rwa}
  B.~Koch, C.~Contreras, P.~Rioseco and F.~Saueressig,
  arXiv:1311.1121.

\bibitem{Emoto:2005te} 
  H.~Emoto,
  hep-th/0511075.

\bibitem{Ward:2006vw} 
  B.~F.~L.~Ward,
  {\it Acta Phys.\ Polon.\ B} {\bf 37}, 1967 (2006),
  hep-ph/0605054.

\bibitem{Falls:2010he} 
  K.~Falls, D.~F.~Litim and A.~Raghuraman,
  {\it Int.\ J.\ Mod.\ Phys.\ A} {\bf 27}, 1250019 (2012),
  arXiv:1002.0260.
  
  
\bibitem{Basu:2010nf} 
  S.~Basu and D.~Mattingly,
  {\it Phys.\ Rev.\ D} {\bf 82}, 124017 (2010);
  arXiv:1006.0718.
  
    \bibitem{Casadio:2010fw}
  R.~Casadio, S.~D.~H.~Hsu and B.~Mirza,
  {\it Phys.\ Lett.\ B} {\bf 695}, 317 (2011);
  arXiv:1008.2768.


\bibitem{Bonanno:2001xi} 
  A.~Bonanno and M.~Reuter,
  Phys.\ Rev.\ D {\bf 65}, 043508 (2002),
  hep-th/0106133.

\bibitem{Bonanno:2001hi} 
  A.~Bonanno and M.~Reuter,
  Phys.\ Lett.\ B {\bf 527}, 9 (2002),
  astro-ph/0106468.

\bibitem{Reuter:2012xf} 
M.\ Reuter and F.\ Saueressig,   {\it Lect.\ Notes Phys.\ } {\bf 863}, 185 (2013),
 arXiv:1205.5431. 
 
\bibitem{Lauscher:2005qz} 
  O.~Lauscher and M.~Reuter,
  JHEP {\bf 0510}, 050 (2005),
  hep-th/0508202.
 
 
\bibitem{Reuter:2011ah} 
  M.~Reuter and F.~Saueressig,
  JHEP {\bf 1112}, 012 (2011),
  arXiv:1110.5224.

\bibitem{Rechenberger:2012pm} 
  S.~Rechenberger and F.~Saueressig,
  {\it Phys.\ Rev.\ D} {\bf 86}, 024018 (2012),
  arXiv:1206.0657.

\bibitem{Calcagni:2013vsa} 
  G.~Calcagni, A.~Eichhorn and F.~Saueressig,
  Phys.\ Rev.\ D {\bf 87}, 124028 (2013),
  arXiv:1304.7247.
  
  
\bibitem{Lauscher:2001ya} 
  O.~Lauscher and M.~Reuter,
  {\it Phys.\ Rev.\ D} {\bf 65}, 025013 (2002),
  hep-th/0108040.
  
\bibitem{Lauscher:2002sq} 
  O.~Lauscher and M.~Reuter,
  {\it Phys.\ Rev.\ D} {\bf 66}, 025026 (2002),
  hep-th/0205062.
  
\bibitem{Manrique:2008zw} 
  E.~Manrique and M.~Reuter,
  Phys.\ Rev.\ D {\bf 79}, 025008 (2009),
  arXiv:0811.3888.

\bibitem{Codello:2013bra} 
  A.~Codello, M.~Demmel and O.~Zanusso,
  arXiv:1310.7625.





\bibitem{Dou:1997fg}
  D.~Dou and R.~Percacci,
  {\it Class.\ Quant.\ Grav.\ } {\bf 15}, 3449 (1998),
  hep-th/9707239.

\bibitem{Souma:1999at}
  W.~Souma,
  {\it Prog.\ Theor.\ Phys.\  } {\bf 102}, 181 (1999),
  hep-th/9907027.

\bibitem{Lauscher:2001rz} 
  O.~Lauscher and M.~Reuter,
  {\it Class.\ Quant.\ Grav.\ }  {\bf 19}, 483 (2002),
  hep-th/0110021.


\bibitem{Reuter:2001ag}
  M.~Reuter and F.~Saueressig,
  {\it Phys.\ Rev.\  D} {\bf 65}, 065016 (2002),
  hep-th/0110054.





\bibitem{Codello:2007bd}
  A.~Codello, R.~Percacci and C.~Rahmede,
  {\it Int.\ J.\ Mod.\ Phys.\  A} {\bf 23}, 143 (2008),
  arXiv:0705.1769.

\bibitem{Machado:2007ea} 
  P.~F.~Machado and F.~Saueressig,
  {\it  Phys.\ Rev.\ D} {\bf 77}, 124045 (2008),
  arXiv:0712.0445.


\bibitem{Benedetti:2009rx} 
  D.~Benedetti, P.~F.~Machado and F.~Saueressig,
  {\it Mod.\ Phys.\ Lett.\ A} {\bf 24}, 2233 (2009),
  arXiv:0901.2984; 
 {\it Nucl.\ Phys.\ B} {\bf 824}, 168 (2010),
  arXiv:0902.4630.
  


\bibitem{Niedermaier:2009zz} 
  M.~R.~Niedermaier,
  {\it Phys.\ Rev.\ Lett.\ } {\bf 103}, 101303 (2009).


\bibitem{Eichhorn:2009ah} 
  A.~Eichhorn, H.~Gies and M.~M.~Scherer,
  {\it Phys.\ Rev.\ D} {\bf 80}, 104003 (2009),
  arXiv:0907.1828;
  K.~Groh and F.~Saueressig,
  {\it J.\ Phys.\ A} {\bf 43}, 365403 (2010),
  arXiv:1001.5032;
  A.~Eichhorn and H.~Gies,
  {\it Phys.\ Rev.\ D} {\bf 81}, 104010 (2010),
  arXiv:1001.5033.
  A.~Eichhorn,
  {\it Phys.\ Rev.\ D} {\bf 87} 124016 (2013), 
  arXiv:1301.0632.


\bibitem{Nagy:2012rn} 
  S.~Nagy, J.~Krizsan and K.~Sailer,
  {\it JHEP} {\bf 1207}, 102 (2012),
  arXiv:1203.6564.
  

\bibitem{Codello:2013wxa} 
  A.~Codello,
  arXiv:1304.2059.
  

\bibitem{Benedetti:2012dx} 
  D.~Benedetti and F.~Caravelli,
  {\it JHEP} {\bf 1206}, 017 (2012)
  [Erratum-ibid.\  {\bf 1210}, 157 (2012)],
  arXiv:1204.3541.
  
\bibitem{Demmel:2012ub} 
  M.~Demmel, F.~Saueressig and O.~Zanusso,
  {\it JHEP} {\bf 1211}, 131 (2012),
  arXiv:1208.2038.
  
\bibitem{Dietz:2012ic} 
  J.~A.~Dietz and T.~R.~Morris,
  {\it JHEP} {\bf 1301}, 108 (2013),
  arXiv:1211.0955.






\bibitem{Reuter:2004nx}
  M.~Reuter and H.~Weyer,
  {\it JCAP} {\bf 0412}, 001 (2004),
  hep-th/0410119.

\bibitem{Christiansen:2012rx} 
  N.~Christiansen, D.~F.~Litim, J.~M.~Pawlowski and A.~Rodigast,
  arXiv:1209.4038.

\bibitem{Litim:2001up} 
  D.~F.~Litim,
  {\it Phys.\ Rev.\ D} {\bf 64}, 105007 (2001),
  hep-th/0103195.




\bibitem{Koch:2010nn} 
  B.~Koch and I.~Ramirez,
   {\it Class.\ Quant.\ Grav.\  } {\bf 28}, 055008 (2011),
  arXiv:1010.2799.
  



\bibitem{Reuter:2003ca}
  M.~Reuter and H.~Weyer,
  {\it Phys.\ Rev.\  D} {\bf 69}, 104022 (2004),
  hep-th/0311196.

\bibitem{Reuter:2005kb} 
  M.~Reuter and F.~Saueressig,
  {\it JCAP} {\bf 0509}, 012 (2005),
  hep-th/0507167.


\bibitem{Domazet:2012tw} 
  S.~Domazet and H.~Stefancic,
  {\it Class.\ Quant.\ Grav.\ }  {\bf 29}, 235005 (2012),
  arXiv:1204.1483.

\bibitem{Frolov:2011ys} 
  A.~V.~Frolov and J.-Q.~Guo,
  arXiv:1101.4995.

\bibitem{Bonanno:2012jy} 
  A.~Bonanno,
  {\it Phys.\ Rev.\ D} {\bf 85}, 081503 (2012),
  arXiv:1203.1962.
  
\bibitem{Hindmarsh:2012rc} 
  M.~Hindmarsh and I.~D.~Saltas,
  {\it Phys.\ Rev.\ D} {\bf 86}, 064029 (2012),
  arXiv:1203.3957.

\bibitem{Copeland:2013vva} 
  E.~J.~Copeland, C.~Rahmede and I.~D.~Saltas,
  arXiv:1311.0881.


\bibitem{Dittrich}
 W.~Dittrich and M.~Reuter, {\it Effective Lagrangians in Quantum Electrodynamics}, (Springer,
Berlin, 1985).


\bibitem{Donoghue:1993eb} 
  J.~F.~Donoghue,
  {\it Phys.\ Rev.\ Lett.\ } {\bf 72}, 2996 (1994),
  gr-qc/9310024.

\bibitem{Larry} L.\ Smarr, {\it Phys.\ Rev.\ Lett.\ } {\bf 30}, 71 (1973).


\bibitem{ArkaniHamed:1998rs}
  N.~Arkani-Hamed, S.~Dimopoulos and G.~R.~Dvali,
  {\it Phys.\ Lett.\  B } {\bf 429}, 263 (1998),
  hep-ph/9803315;
  I.~Antoniadis, N.~Arkani-Hamed, S.~Dimopoulos and G.~R.~Dvali,
  {\it Phys.\ Lett.\  B } {\bf 436}, 257 (1998),
  hep-ph/9804398;
  N.~Arkani-Hamed, S.~Dimopoulos and G.~R.~Dvali,
  {\it Phys.\ Rev.\  D } {\bf 59}, 086004 (1999),
  hep-ph/9807344.

\bibitem{Randall:1999vf}
  L.~Randall and R.~Sundrum,
  {\it Phys.\ Rev.\ Lett.\ }  {\bf 83}, 4690 (1999),
  hep-th/9906064;
  L.~Randall and R.~Sundrum,
  {\it Phys.\ Rev.\ Lett.\ } {\bf 83}, 3370 (1999),
  hep-ph/9905221.

\bibitem{Banks:1999gd}
  T.~Banks and W.~Fischler,
  hep-th/9906038.

\bibitem{Giddings:2000ay}
  S.~B.~Giddings and E.~Katz,
  {\it J.\ Math.\ Phys.\ }  {\bf 42}, 3082 (2001),
  hep-th/0009176.

\bibitem{Giddings:2001bu}
  S.~B.~Giddings and S.~D.~Thomas,
  {\it Phys.\ Rev.\  D} {\bf 65}, 056010 (2002),
  hep-ph/0106219.

\bibitem{Dimopoulos:2001hw}
  S.~Dimopoulos and G.~L.~Landsberg,
  {\it Phys.\ Rev.\ Lett.\ }  {\bf 87}, 161602 (2001),
  hep-ph/0106295.

\bibitem{Hossenfelder:2001dn}
  S.~Hossenfelder, S.~Hofmann, M.~Bleicher and H.~Stoecker,
  {\it Phys.\ Rev.\  D } {\bf 66}, 101502 (2002),
  hep-ph/0109085.

\bibitem{Bleicher:2001kh}
  M.~Bleicher, S.~Hofmann, S.~Hossenfelder and H.~Stoecker,
  {\it Phys.\ Lett.\  B } {\bf 548}, 73 (2002),
  hep-ph/0112186.

\bibitem{Fischer:2006fz} 
  P.~Fischer and D.~F.~Litim,
 {\it Phys.\ Lett.\ B} {\bf 638}, 497 (2006),
  hep-th/0602203.
  
 
\bibitem{Koch:2007yt}
  B.~Koch,
  {\it Phys.\ Lett.\  B} {\bf 663}, 334 (2008);
  arXiv:0707.4644.

\bibitem{Hewett:2007st} 
  J.~Hewett and T.~Rizzo,
  {\it JHEP} {\bf 0712}, 009 (2007),
  arXiv:0707.3182.

\bibitem{Litim:2007iu} 
  D.~F.~Litim and T.~Plehn,
  {\it Phys.\ Rev.\ Lett.\ }  {\bf 100}, 131301 (2008),
  arXiv:0707.3983.

\bibitem{Burschil:2009va}
  T.~Burschil and B.~Koch,
  {\it Zh.\ Eksp.\ Teor.\ Fiz.\  } {\bf 92}, 219 (2010),
  arXiv:0912.4517.

\bibitem{Contreras:2013hua} 
  C.~Contreras, B.~Koch and P.~Rioseco,
  arXiv:1303.3892.
 
\bibitem{Koch:2005ks} 
  B.~Koch, M.~Bleicher and S.~Hossenfelder,
  JHEP {\bf 0510}, 053 (2005)
  [hep-ph/0507138].

\bibitem{Bellagamba:2012wz} 
  L.~Bellagamba, R.~Casadio, R.~Di Sipio and V.~Viventi,
  Eur.\ Phys.\ J.\ C {\bf 72}, 1957 (2012)
  [arXiv:1201.3208 [hep-ph]].
  
\bibitem{Alberghi:2013hca} 
  G.~L.~Alberghi, L.~Bellagamba, X.~Calmet, R.~Casadio and O.~Micu,
  Eur.\ Phys.\ J.\ C {\bf 73}, 2448 (2013)
  [arXiv:1303.3150 [hep-ph]].

\bibitem{Voloshin:2001fe}
  M.~B.~Voloshin,
  {\it Phys.\ Lett.\  B} {\bf 524}, 376 (2002)
  [Erratum-ibid.\  B {\bf 605}, 426 (2005)],
  hep-ph/0111099;
  M.~B.~Voloshin,
  {\it Phys.\ Lett.\  B} {\bf 518}, 137 (2001),
  hep-ph/0107119.

\bibitem{Jevicki:2002fq}
  A.~Jevicki and J.~Thaler,
  {\it Phys.\ Rev.\  D} {\bf 66}, 024041 (2002),
  hep-th/0203172.

\bibitem{Eardley:2002re}
  D.~M.~Eardley and S.~B.~Giddings,
  {\it Phys.\ Rev.\  D} {\bf 66}, 044011 (2002),
  gr-qc/0201034.

\bibitem{Rychkov:2004sf}
  V.~S.~Rychkov,
  {\it Phys.\ Rev.\  D} {\bf 70}, 044003 (2004),
  hep-ph/0401116.

\bibitem{Kang:2004yk}
  K.~Kang and H.~Nastase,
  {\it Phys.\ Rev.\  D} {\bf 71}, 124035 (2005),
  hep-th/0409099.
  
\bibitem{Rizzo:2006uz}
  P.~Nicolini,
  {\it J.\ Phys.\ A } {\bf 38}, L631 (2005)
  hep-th/0507266;
  T.~G.~Rizzo,
  {\it Class.\ Quant.\ Grav.\ }  {\bf 23}, 4263 (2006),
  hep-ph/0601029;
  T.~G.~Rizzo,
  {\it JHEP} {\bf 0609}, 021 (2006),
  hep-ph/0606051;
  E.~Spallucci, A.~Smailagic and P.~Nicolini,
  {\it Phys.\ Lett.\  B} {\bf 670}, 449 (2009),
  arXiv:0801.3519.
  
  
\bibitem{Carlip:2009kf}
  S.~Carlip,
  arXiv:0909.3329.

\bibitem{Futamase:1984ej} 
  T.~Futamase,
  {\it Phys.\ Rev.\ D} {\bf 29}, 2783 (1984).

\bibitem{Hu:1986cv} 
  B.~L.~Hu and D.~J.~O'Connor,
  {\it Phys.\ Rev.\ D} {\bf 36}, 1701 (1987).

\bibitem{Berkin:1992ub} 
  A.~L.~Berkin,
  {\it Phys.\ Rev.\ D} {\bf 46}, 1551 (1992).

\bibitem{Benedetti:2010nr} 
  D.~Benedetti, K.~Groh, P.~F.~Machado and F.~Saueressig,
  {\it JHEP} {\bf 1106}, 079 (2011),
  arXiv:1012.3081.

\bibitem{Manrique:2010mq}
  E.~Manrique, M.~Reuter and F.~Saueressig,
  {\it Annals Phys.\ } {\bf 326}, 440 (2011),
  arXiv:1003.5129.

\bibitem{Manrique:2010am}
  E.~Manrique, M.~Reuter and F.~Saueressig,
  {\it Annals Phys.\ }  {\bf 326}, 463 (2011),
  arXiv:1006.0099.
  
  

\bibitem{Julve:1978xn}
  J.~Julve and M.~Tonin,
  Nuovo Cim.\ B {\bf 46}, 137 (1978).

\bibitem{Salam:1978fd}
  A.~Salam and J.~A.~Strathdee,
  {\it Phys.\ Rev.\ D } {\bf 18}, 4480 (1978).

\bibitem{Codello:2006in}
  A.~Codello and R.~Percacci,
  {\it Phys.\ Rev.\ Lett.\ } {\bf 97}, 221301 (2006),
  hep-th/0607128.

\bibitem{Groh:2011vn}
  K.~Groh, S.~Rechenberger, F.~Saueressig and O.~Zanusso,
  PoS EPS {\bf -HEP2011}, 124 (2011),
  arXiv:1111.1743.
  
  
\bibitem{Andrade:2006pg} 
  T.~Andrade, M.~Banados and F.~Rojas,
  {\it Phys.\ Rev.\ D} {\bf 75}, 065013 (2007),
  hep-th/0612150.


\end{thebibliography}
\end{document}